\documentclass[aps, prb, twocolumn, amssymb, amsmath, showpacs, superscriptaddress]{revtex4-1}
\usepackage{graphicx}
\usepackage{epstopdf}
\usepackage{dcolumn}
\usepackage{bm}
\usepackage{times}
\usepackage{booktabs}
\usepackage{setspace}
\usepackage[colorlinks,
            linkcolor=red,
            anchorcolor=green,
            citecolor=blue
            ]{hyperref}

\newcommand{\s}{{\sigma}}

\def\eqa{\begin{eqnarray}}
\def\eea{\end{eqnarray}}
\newcommand{\eq}{\begin{equation}}
\newcommand{\ee}{\end{equation}}
\newcommand{\nn}{\nonumber\\}

\begin{document}

\title {Triplet $p_z$-wave pairing in quasi one dimensional A$_2$Cr$_3$As$_3$ superconductors}

\author{Xianxin Wu}
\affiliation{ Institute of Physics, Chinese Academy of Sciences,
Beijing 100190, China}

\author{Fan Yang}\email{yangfan\_blg@bit.edu.cn}
\affiliation{School of Physics, Beijing Institute of Technology, Beijing, 100081, China}

\author{Congcong Le }  \affiliation{ Institute of Physics, Chinese Academy of Sciences,
Beijing 100190, China}

\author{Heng Fan}  \affiliation{ Institute of Physics, Chinese Academy of Sciences,
Beijing 100190, China} \affiliation{Collaborative Innovation Center of Quantum Matter, Beijing, China}

\author{Jiangping Hu  }\email{jphu@iphy.ac.cn} \affiliation{ Institute of Physics, Chinese Academy of Sciences,
Beijing 100190, China}\affiliation{Department of Physics, Purdue University, West Lafayette, Indiana 47907, USA}
\affiliation{Collaborative Innovation Center of Quantum Matter, Beijing, China}

\date{\today}

\begin{abstract}
We construct  minimum effective models to investigate the pairing symmetry in  the newly discovered quasi-one-dimensional superconductor K$_2$Cr$_3$As$_3$.  We show that a minimum three-band model based on the $d_{z^2}$, $d_{xy}$ and $d_{x^2-y^2}$ orbitals of one Cr sublattice can  capture the band structures near Fermi surfaces. In both weak and strong coupling limits,  the standard random phase approximation (RPA) and mean-field solutions consistently yield the triplet $p_z$-wave pairing as the leading pairing symmetry for physically realistic parameters. The triplet pairing is driven by the ferromagnetic fluctuations within the sublattice. The gap function of the pairing state possesses line gap nodes on the $k_z=0$ plane on the Fermi surfaces.
\end{abstract}


\maketitle

\section {Introduction:}
Searching for triplet superconductivity (SC) has been  one of  major research efforts recently partly due to its  intrinsic connection to topologically related physics and quantum computations.   Unconventional triplet SC  is normally driven by electron-electron interactions.  Until now, the candidates for triplet pairing  superconductors are rather limited. For  quasi-two dimensional electron systems,  Sr$_2$RuO$_4$\cite{SrRuO1,SrRuO2} is considered to be a possible good candidate. For quasi-one-dimensional(Q1D)  systems,   since the electron-electron correlation can be further enhanced,  triplet pairing has also been proposed in previously discovered Q1D systems, such as Bechgaard salts\cite{Jerome1980,Wilhelm2001} , Tl$_2$Mo$_6$Se$_6$\cite{Armici1980} and  Li$_{0.9}$Mo$_6$O$_{17}$\cite{Greenblatt1984,Denlinger1999,Xu2009,Mercure2012}.  However, in these Q1D materials, the correlation may be offset by the fact  that their electronic physics is not attributed to  the $3d$ orbitals which are known to produce the strongest correlation effect due to small orbital size.  Thus, the newly synthesized Q1D superconductors A$_2$Cr$_3$As$_3$(A=K,Rb,Cs)\cite{KCrAs,RbCrAs,CsCrAs} become intriguing materials to explore possible triplet pairing because their electronic structures are attributed to the $3d$ orbitals of chromium atoms.

The chromium-based superconductors have attracted great research interests  recently\cite{Luo,Kotegawa,KCrAs,RbCrAs,CsCrAs}.  Their superconducting mechanism shown in many experimental results\cite{Luo,Kotegawa,Zhaojun, NMR_CrAs}  is very likely  related to  electron-electron interactions. Among the families of the chromium-based superconductors,  the Q1D A$_2$Cr$_3$As$_3$ (A=K,Rb,Cs) superconductors family\cite{KCrAs,RbCrAs,CsCrAs}, which has $T_c$ up to 6.1 K\cite{KCrAs} at ambient pressure, is particularly intriguing.  The crystal of this family consists of alkali metal ions separated [(Cr$_3$As$_3$)$^{2-}$]$_{\infty}$ double-walled subnanotubes.  The electronic structures calculated from  Density Functional Theory (DFT)  suggests that the $3d$ orbitals of Cr make  major contribution near Fermi surfaces(FSs) \cite{Cao,Wu1}.  The large electronic specific-heat coefficient\cite{KCrAs},  the non-fermi liquid behavior in the normal state \cite{KCrAs} and the abnormality in NMR relaxation-rate\cite{NMR_KCrAs}  all suggest  possible strong electron correlations in the family. The linearly temperature-dependent London penetration depth\cite{pen_depth}, the lack of the Hebel-Slichter peak in NMR relaxation-rate upon the superconducting transition\cite{NMR_KCrAs}, and the $\sqrt{H}$-dependence of $C_v$ in the mixed state\cite{RbCrAs} suggest unconventional SC with line nodes in the system. More importantly, the upper critical field largely exceeds the Pauli limit\cite{KCrAs,RbCrAs,CsCrAs}, implying possible triplet pairings in the family.

The triplet pairings were also argued theoretically\cite{Wu1}. The theoretical calculations predicted that the materials are very close to a new in-out coplanar (IOP) magnetic state.  The magnetic state is caused by  the strong antiferromagnetic couplings between the nearest neighbour (NN) Cr atoms and the weak ferromagnetic couplings between the next nearest neighbour (NNN) ones.  Intriguingly, the magnetic fluctuations associated with such a magnetic state  can favor  triplet pairing.   A unit cell of A$_2$Cr$_3$As$_3$ is composed of two types of inequivalent Cr atoms which will be referred as Cr1 and Cr2 atoms as shown in Fig.\ref{model}, which form two sublattices, A and B, respectively.     It is shown that Q1D  electronic structures   are dominantly contributed  by only the Cr2 sublattice\cite{Wu1}.  Within the Cr2 sublattice, the magnetic fluctuations are ferromagnetic(FM), which are caused by both the weak NNN ferromagnetic couplings and effective ferromagnetic fluctuations generated by the NN antiferromagnetism. Thus the triplet pairing is very possible in A$_2$Cr$_3$As$_3$.  Furthermore,  the materials do not have space inversion symmetry but it still has the mirror plane symmetry with respect to  the $x-y$ plane. Therefore, along the Q1D chain direction, the  triplet $p_z$-wave pairing, which is odd with respect to the mirror plane reflection,  is a natural triplet candidate classified by the irreducible representations of the lattice symmetry.

In this paper, we construct low energy effective models to study the pairing symmetry of K$_2$Cr$_3$As$_3$. First, we construct the minimum tight binding model for the material.   A six-band model is constructed with the Cr-$d_{z^2}$, $d_{xy}$ and $d_{x^2-y^2}$ orbitals of both Cr1 and Cr2 atoms. At low energy, this model can be further simplified to a three-band model with only the $d$-orbitals  of the Cr2 atoms.  Starting from the three-band tight-binding model,  we investigate the pairing symmetries in two types of standard  models, one with the standard electron-electron interaction terms including onsite Hubbard $U$ and Hund's coupling $J_H$ and the other with the effective FM exchange couplings.  We adopt combined random phase approximation (RPA) and mean-field approaches for the weak and strong interaction limits respectively. Both approaches consistently yield the $p_z$-wave triplet pairing as the leading pairing symmetry for physically realistic parameters satisfying $\frac{J_H}{U}<r_c\approx\frac{1}{3}$ (for $\frac{J_H}{U}>r_c$, the RPA yields on-site inter-orbital $f_{y^3-3x^2y}$-wave pairing, driven by the Hund's rule coupling), driven by intra-sublattice FM spin fluctuations. The gap function of this pairing possesses line gap nodes on the $k_z=0$ plane on the FS. This pairing is essentially Q1D and the pairing amplitudes on the Q1D $\beta$ band dominates those on the Q1D $\alpha$ band and the 3D $\gamma$- band.  Finally, we shall address the experimental relevance of the $p_z$-wave triplet pairing obtained here.

\section{Model and approach}
\begin{figure}[tb]
\centerline{\includegraphics[height=8 cm]{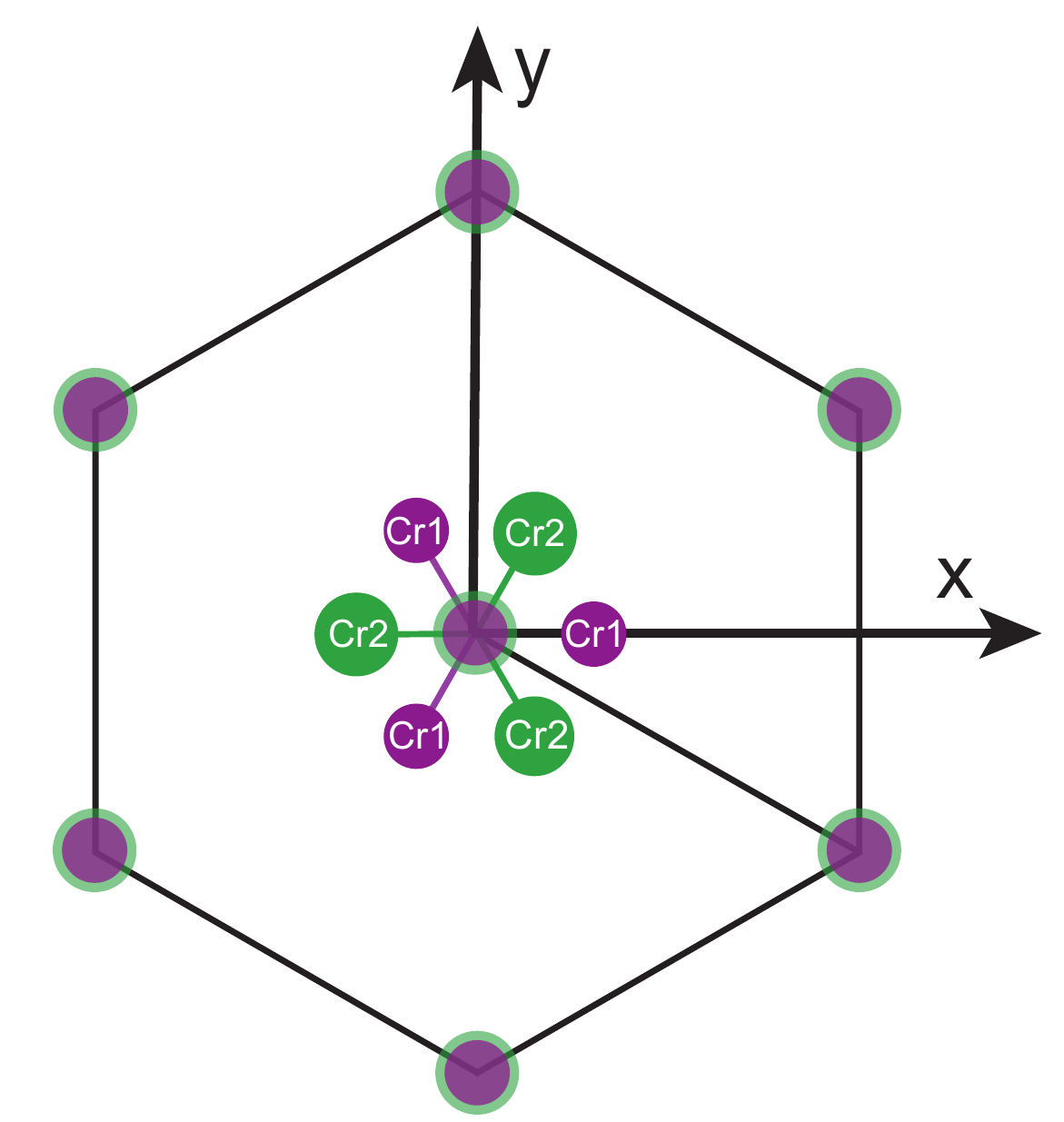}}
\caption{(color online). The lattice model with virtual atoms. The purple and green circles on each lattice site represent the virtual atoms of Cr1 and Cr2, respectively.  \label{model} }
\end{figure}

%\subsection {The tight-binding Model}
From the bandstructure of DFT calculations\cite{Cao,Wu1}, it's found that there are five bands near the Fermi level(FL) and the oribtal characters near the Fermi level are mainly of Cr- $d_{z^2}$ ($A'_{1}$ symmetry) and ($d_{xy}$, $d_{x^2-y^2}$) ($E'$ symmetry) components. To obtain an effective model, we consider a virtual atom at the center of each Cr triangle with three orbitals, which have the same symmetries as $A'_1$ and $E'$, as shown in Fig.\ref{model}. Now the structure of K$_2$Cr$_3$As$_3$ is simplified as parallel placed Cr-chains, with each chain containing  the A and B sublattices,  which are inequivalent due to the asymmetry of K atoms and the absence of inversion symmetry. The point group of this new lattice model is the same with the original lattice and it is $D_{3h}$ not $D_{6h}$, where inversion and six-fold rotational operations are absent. The tight-binding Hamiltonian thus obtained in the momentum space can be written as,
 \begin{eqnarray}
 H^{(0)}_{\rm TB}=\sum_{\mathbf{k}mn\mu\nu\s}h^{mn}_{\mu\nu}(\mathbf{k})c^\dag_{m\mu\sigma}(\mathbf{k})c_{n\nu\sigma}(\mathbf{k}),
 \end{eqnarray}
 where $m/n=A,B$ labels sublattice, $\mu$/$\nu=1,\cdots,3$ labels orbital and $\sigma=\uparrow,\downarrow$ labels spin. $c^\dag_{m\mu\sigma}(\mathbf{k})$ creates a spin-$\sigma$ electron in the orbital $\mu$ in the $m$-th sublattice with momentum $\mathbf{k}$. The elements of the matrix $h^{mn}_{\mu\nu}(\mathbf{k})$ and the resulting band structure of this model are given in Appendix.A. The band structure of this six-band tight-binding model well captures the main characters of that of DFT at low energy.
\begin{figure}
\scalebox{0.5}{\includegraphics[scale=0.6]{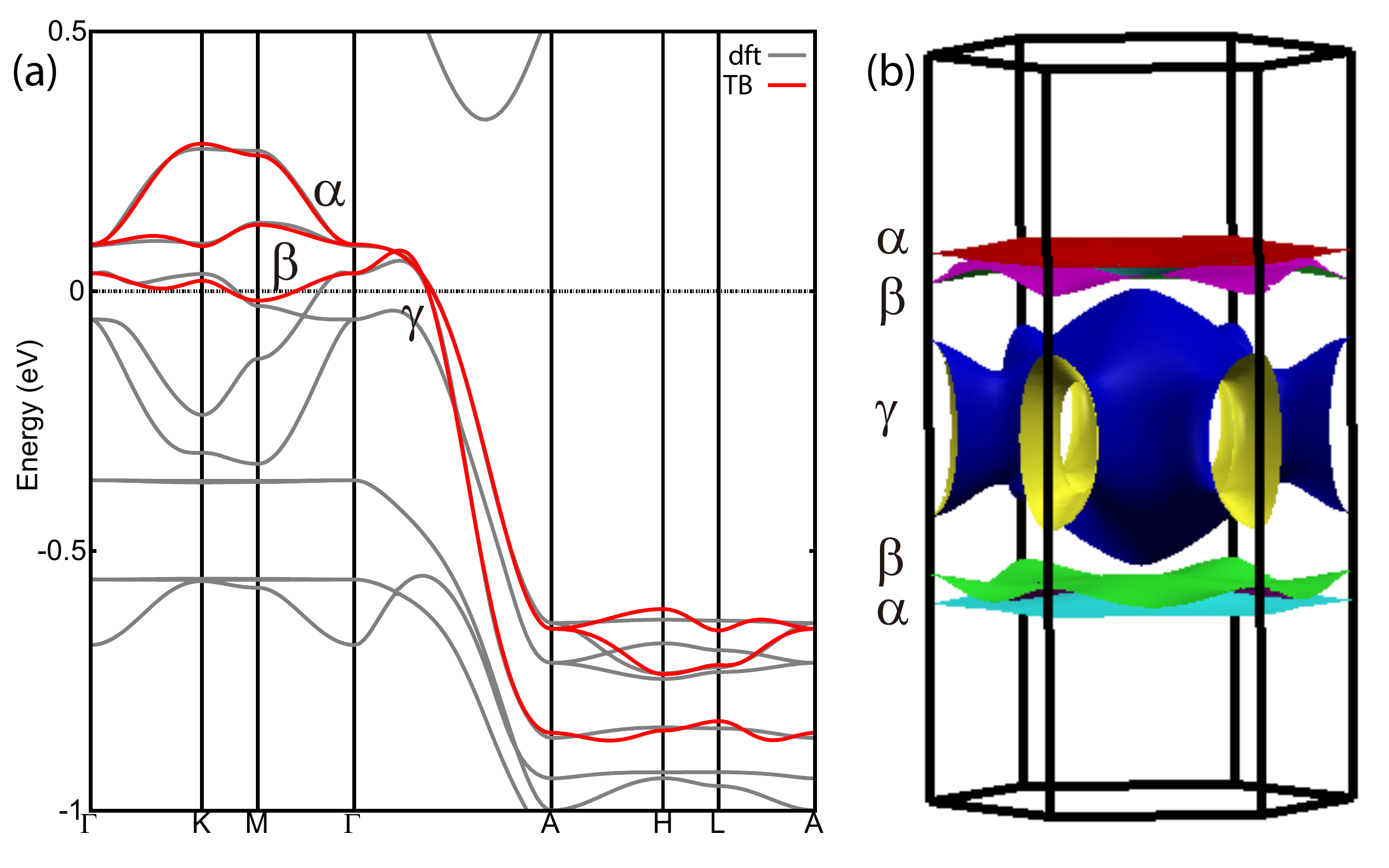}}
\caption{(color online). Band structure and Fermi surfaces. (a).The band structures of DFT (gray lines) and three-band tight-binding model(red lines). The $\alpha$- and $\beta$-bands are Q1D and the $\gamma$-band is 3D. (b). FSs of the tight-binding model.} \label{bandTB3}
\end{figure}
\begin{figure}
\scalebox{0.32}{\includegraphics[scale=0.5]{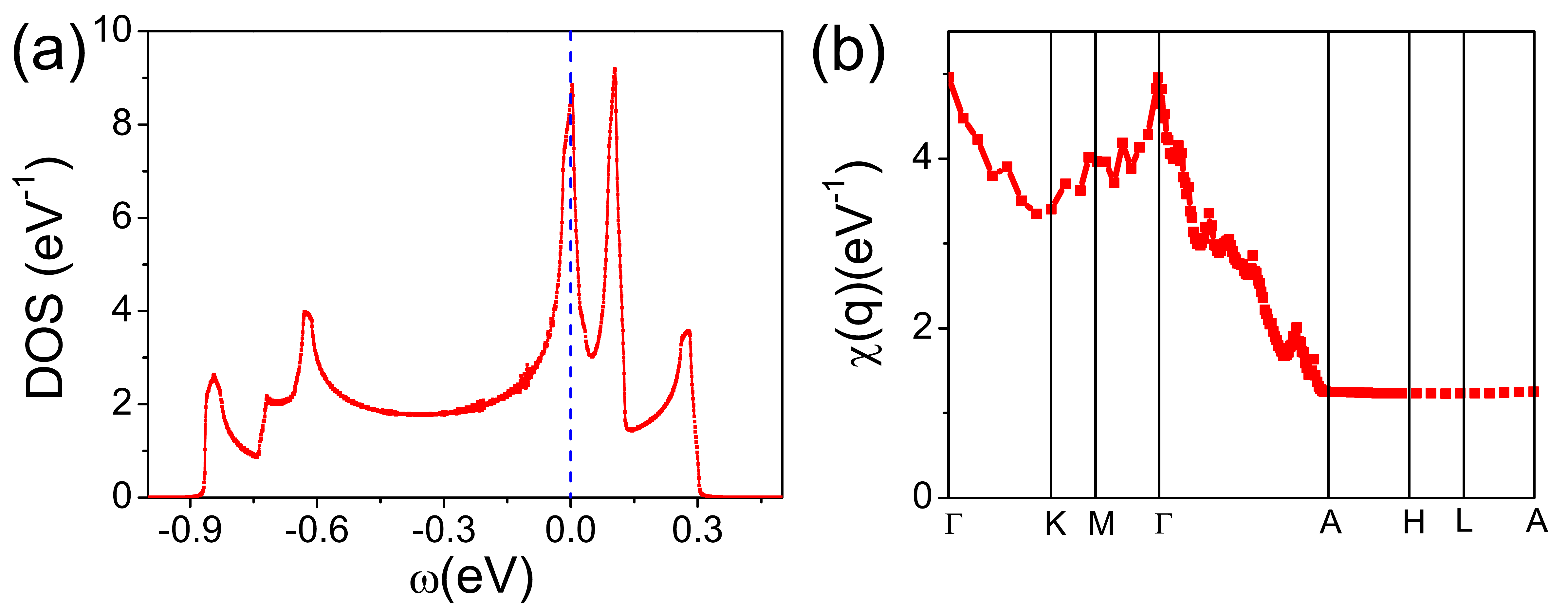}
}\caption{(color online). Density of states and susceptibility. (a). The DOS (for one spin specie) of the three-band tight-binding model. (b). Momentum-dependence of the eigen susceptibility $\chi(\mathbf{q})$ along the high-symmetry lines in the Brilloiun-Zone. The temperature is $T=0.001$eV.} \label{DOS_sus}
\end{figure}

\begin{table*}[bt]%The best place to locate the table environment is directly after its first reference in text
\caption{\label{hopping3band} The hopping parameters  in  the three-orbital tight-binding model. The onsite energies of $A'_{1}$ and $E'$ orbitals are $\epsilon_1=2.0939$ and $\epsilon_2=2.0997$. They are given in unit of eV. }
\begin{ruledtabular}
\begin{tabular}{ccccccccccc}
%\textrm{Left\footnote{Note a.}}& \textrm{Centered\footnote{Note
%b.}}& \multicolumn{1}{c}{\textrm{Decimal}}&
%\textrm{Right}\\
$s^{\alpha\beta}_i$ &  $i=z,2$  & $i=z,4$  & $i=z,6$ &  $i=1y$ & $i=2y$ & $i=11y$ & $i=22y$ & $i=12y$ & $i=y$ & $i=yz$ \\
 \colrule
$\alpha\beta$=11   & 0.2841  & -0.0505  & -0.0252 &    &   &    &    &   & -0.0151 &  -0.0063    \\
$\alpha\beta$=22   & 0.1816 & -0.0469   &  0.0034 &          &     & 0.0216  &-0.0135  & 0.0064 & &    \\
$\alpha\beta$=12   &        &           &         & 0.0175  & 0.0095 &  &   &   &  &        \\
%$mn$=22,$s^{22}_{i}$&          &     & 0.0216  &-0.0135  & 0.0064 & &           \\
%$mn$=11,$s^{11}_{i}$  &    &   &    &    &   & -0.0151 &  -0.0063  \\
\end{tabular}
\end{ruledtabular}
\end{table*}

Notice that only three of the six bands cross the Fermi surface, which provides possibility for us to further simplify our model to a three-band one. This is accomplished by hybridizing each d-orbital from Cr1 and that from Cr2 into a lower energy d-orbital, and we finally arrive at a three-orbital model. Neglecting spin-orbit coupling, this model is described by the following tight-binding Hamiltonian
\begin{equation}
H_{\rm TB}=\sum_{\mathbf{k}\mu\nu\s}h_{\mu\nu}(\mathbf{k})c^{\dagger}_{\mu\s}(\mathbf{k})c_{\nu\s}(\mathbf{k}).\label{TB}
\end{equation}
Here, the orbital index $\mu/\nu=1,\cdots,3$ represent the $d_{z^2}$ for 1, the $d_{xy}$ for 2, and the $d_{x^2-y^2}$ for 3, respectively. The matrix elements $h_{\mu\nu}(\mathbf{k})$ are given by
\begin{eqnarray}
h_{11}&=&\epsilon_{1}+(s^{11}_y+2s^{11}_{yz}{\rm cos} 2z)(2{\rm cos}2y+4{\rm cos} x {\rm cos} y )\nonumber\\&&+2\sum_{j=1}^{3}s^{11}_{z,2j}{\rm cos}2jz,\nonumber\\
h_{12}&=&2is^{12}_{1y}{\rm sin} 2y +2is^{12}_{1y}{\rm sin} y {\rm cos} x +2\sqrt{3}s^{12}_{2y}{\rm sin} y {\rm sin} x,\nonumber\\
h_{21}&=&h_{12}^{*},\nonumber\\
h_{13}&=&2s^{12}_{2y}{\rm cos} 2y -2i\sqrt{3}s^{12}_{1y}{\rm cos} y {\rm sin} x -2s^{12}_{2y}{\rm cos} y {\rm cos} x,\nonumber\\
h_{31}&=&h_{13}^{*},\nonumber\\
h_{22}&=&\epsilon_{2}+2s^{22}_{11y}{\rm cos} 2y+(s^{22}_{11y}+3s^{22}_{22y}){\rm cos} x {\rm cos} y \nonumber\\&&+2\sum_{j=1}^{2}s^{22}_{z,2j}{\rm cos}2jz,\nonumber\\
h_{23}&=&\sqrt{3}(s^{22}_{11y}-s^{22}_{22y}){\rm sin} x {\rm sin} y +2is^{22}_{12y}{\rm sin} 2y ,\nonumber\\&&-4is^{22}_{12y}{\rm cos} x {\rm sin}y,\nonumber\\
h_{32}&=&h_{23}^{*},\nonumber\\
h_{33}&=&\epsilon_{2}+2s^{22}_{22y}{\rm cos} 2y+(3s^{22}_{11y}+s^{22}_{22y}){\rm cos} x {\rm cos} y, \nonumber\\&&+2\sum_{j=1}^{2}s^{22}_{z,2j}{\rm cos}2jz.\label{matrix}
\end{eqnarray}
Here $x=\frac{\sqrt{3}}{2}{k_xa_0}$, $y=\frac{1}{2}k_ya_0$ and $z=\frac{1}{2}k_zc_0$, where $a_0,c_0$ are lattice constants within the $xy$- plane and along the $z$-axis respectively. The tight-binding parameters $\epsilon_{1,2}$ and $s^{\alpha\beta}_{i}$ are shown in Table \ref{hopping3band}.

The band structure of the above introduced three-band tight-binding model is shown in Fig.\ref{bandTB3}(a), in comparison with that of DFT. The chemical potential of our tight-binding model is $\mu_c=2.3105$eV, leading to a band filling of 4-electrons per unit cell. From the comparison, it's clear that the three-band tight-binding band structure well captures the main characters of the DFT band structure at low energy. Two 1D bands (marked as $\alpha$ and $\beta$ in Fig.\ref{bandTB3} (a)) and one 3D band (marked as $\gamma$) are found near the FL. The two 1D bands are degenerate along the $\Gamma-A$ line. The FSs of the tight-binding model are shown in Fig.\ref{bandTB3}(b), where the FS sheets are marked corresponding to the three bands. The dominating orbital component on the $\alpha$- FS is $d_{z^2}$, while those on the $\beta$- and $\gamma$- FSs are $(d_{xy}, d_{x^2-y^2})$ (see Appendix.A).  Due to the flatness of the two 1D bands near the FS as shown in Fig.\ref{bandTB3}(b), large density of states (DOS) is obtained at the Fermi level, as shown in Fig.\ref{DOS_sus}(a).

To generate superconductivity, interactions are necessary. Here we focus on the possible pairing mechanism driven by electron-electron interactions. Since the interaction strength of the $3d$-electrons of Cr might be moderate, we shall investigate both the weak-coupling and strong-coupling limits of the electron interactions in the following, with both limits giving consistent results. For the weak coupling limits, we adopted the Hubbard-Hund's type of on-site interactions and performed the standard multi-orbital RPA\cite{RPA1,RPA2,RPA3,Kuroki,Scalapino1,Scalapino2,Yang,Wu2014,Ma2014,Zhang2015} calculations to study the pairing symmetry. For the strong coupling limit, we adopted the super-exchange type of interactions and performed a mean-field analysis on the $t-J$ type of model instead.

\section{The weak coupling limit}
%\subsection{Interactions and Approaches}

In the weak coupling limit, the following on-site Hubbard-Hund's interactions are adopted
\begin{eqnarray}
H_{\rm H-H}&=&U\sum_{i\mu}n_{i\mu\uparrow}n_{i\mu\downarrow}+
V\sum_{i,\mu<\nu}n_{i\mu}n_{i\nu}+J_{H}\sum_{i,\mu<\nu}\Big[\nn&&
\sum_{\sigma\sigma^{\prime}}c^{\dag}_{i\mu\sigma}c^{\dag}_{i\nu\sigma^{\prime}}
c_{i\mu\sigma^{\prime}}c_{i\nu\sigma}+(c^{\dag}_{i\mu\uparrow}c^{\dag}_{i\mu\downarrow}
c_{i\nu\downarrow}c_{i\nu\uparrow}+h.c.)\Big].\nonumber\\\label{H-H-model}
\end{eqnarray}
Here, the $U$-term, the $V$-term and the $J_H$-term denote the intra-orbital, inter-orbital Hubbard-repulsion and the Hund's rule coupling as well as the pair hopping. For simplicity, we have assumed that the coupling constants $U, V, J_H$ do not depend on the orbital indices in our effective model. General symmetry argument requires $U=V+2J_H$. Note that physically the orbital bases $d_{z^2}$,$d_{xy}$ and $d_{x^2-y^2}$ building our effective model are not the maximumly-localized Wannier wave functions around a real single Cr-atom, but the mixture of the three atomic wave functions around the virtual center. Therefore, the on-site interactions on these delocalized orbital bases are much weaker than that on a real single atom. Furthermore, the interaction parameters $U,V$ and $J_H$ should be expressed by those interaction parameters on the local Wannier basis, which in general can be estimated from first principle approaches such as cRPA, although such estimation is usually coarse and strongly approach-dependent. Here, we let $U$ and the ratio $r\equiv J_H/U$ to be tuning parameters and study the parameter-dependence of the results. We left the determination of the interaction parameters for future study.

To study the pairing symmetry of this Hubbard-Hund model, we adopt the standard multi-orbital RPA approach\cite{RPA1,RPA2,RPA3,Kuroki,Scalapino1,Scalapino2,Yang,Wu2014,Ma2014,Zhang2015}. We first define and calculate the bare susceptibility tensor $\chi^{(0)l_{1},l_{2}}_{l_{3},l_{4}}\left(\mathbf{q},\tau\right)$. After that, the renormalized spin(s) or charge(c) susceptibilities $\chi^{(s,c)l_{1},l_{2}}_{l_{3},l_{4}}\left(\mathbf{q},\tau\right)$ are obtained in the RPA level. Then, through exchanging the spin or charge fluctuations, we obtain the effective pairing potential $V^{\alpha\beta}(\mathbf{k,q})$. Finally, solving the linearized gap equation for $V^{\alpha\beta}(\mathbf{k,q})$ as an eigenvalue problem, we obtain the leading pairing gap function as the eigenvector corresponding to the largest eigenvalue $\lambda$. Generally, the RPA only works well for weak interactions.

%\section{The weak coupling limit}
For the weak-coupling limit, we adopt multi-orbital RPA approach\cite{RPA1,RPA2,RPA3,Kuroki,Scalapino1,Scalapino2,Yang,Wu2014,Ma2014,Zhang2015} to study the pairing symmetry of the Hubbard-Hund's model of the system. The Hamiltonian of our model is,
\begin{eqnarray}
H=H_{\rm TB}+H_{\rm H-H},\label{H_H_H}
\end{eqnarray}
where the interaction term $H_{\rm H-H}$ is give by Eq.(\ref{H-H-model}).

\subsection {Susceptibilities and magnetic phase}
 Let's define the following bare susceptibility tensor,
  \begin{eqnarray}
 \chi^{(0)l_{1},l_{2}}_{l_{3},l_{4}}\left(\mathbf{q},\tau\right)\equiv
 \frac{1}{N}\sum_{\mathbf{k_{1},k_{2}}}&&\left<T_{\tau}c^{\dagger}_{l_{1}}(\mathbf{k_{1}},\tau)
 c_{l_{2}}(\mathbf{k_{1}+q},\tau)\right.\nonumber\\&&\left.c^{+}_{l_{3}}(\mathbf{k_{2}+q},0)c_{l_{4}}(\mathbf{k_{2}},0)\right>_0,\label{chi_0}
 \end{eqnarray}
  with $l_{i(i=1,\cdots,4)}$ denoting orbital indices. The explicit formalism of $\chi^{(0)}$ is given by
\begin{eqnarray}
 \chi^{(0)l_{1},l_{2}}_{l_{3},l_{4}}\left(\mathbf{q},i\omega_n\right)&=&\frac{1}{N}\sum_{\mathbf{k},\alpha,\beta}
 \xi^{\alpha}_{l_{4}}(\mathbf{k})\xi_{l_{1}}^{\alpha,*}(\mathbf{k})\xi^{\beta}_{l_{2}}(\mathbf{k+q})\nonumber\\&&\xi^{\beta,*}_{l_{3}}(\mathbf{k+q})
 \frac{n_{F}(\varepsilon^{\beta}_{\mathbf{k+q}})-n_{F}(\varepsilon^{\alpha}_{\mathbf{k}})}{i\omega_n+\varepsilon^{\alpha}_{\mathbf{k}}-\varepsilon^{\beta}_{\mathbf{k+q}}},\label{explicit_free}
 \end{eqnarray}
 where $\alpha/\beta=1,...,3$ are band indices, $\varepsilon^{\alpha}_{\mathbf{k}}$ and $\xi^{\alpha}_{l}\left(\mathbf{k}\right)$ are the $\alpha-$th eigenvalue and eigenvector of the $h(\mathbf{k})$ matrix respectively and $n_F$ is the Fermi-Dirac distribution function.

To detect the pattern of the dominating spin correlation in the system, we define the static susceptibility matrix $\chi_{\mathbf{q}}(l,m)\equiv \chi^{(0)l,l}_{m,m}(\mathbf{q},i\omega_n=0)$. The largest eigenvalue of this Hermitian matrix for each momentum $\mathbf{q}$ is defined as $\chi(\mathbf{q})$, which represents the eigen susceptibility in the strongest channel. The distribution of $\chi(\mathbf{q})$ in the Brillouin-Zone reflects the pattern of spin correlations in the system. Fig.\ref{DOS_sus}(b) shows the momentum-dependence of $\chi(\mathbf{q})$ along the high-symmetry lines, where the peak at the $\Gamma$-point reveals strong FM spin correlations in the model. Note that in the building of our model, only one sublattice is taken from real material, which thus means that the spin correlation pattern found here is intra-sublattice FM.

When interactions turn on, the renormalized spin(s) or charge(c) susceptibilities are obtained in the RPA level as
 \begin{eqnarray}
\chi^{(s,c)}\left(\mathbf{q},i\nu\right)&=&\left[I\mp\chi^{(0)}\left(\mathbf{q},i\nu\right)U^{(s,c)}\right]^{-1}\chi^{(0)}\left(\mathbf{q},i\nu\right).\label{RPA_SUS}
\end{eqnarray}
Here $\chi^{(s,c)}\left(\mathbf{q},i\nu_{n}\right)$, $\chi^{(0)}\left(\mathbf{q},i\nu_{n}\right)$ and $U^{(s,c)}$ are operated as
$9\times 9$ matrices (the upper or lower two indices are viewed as one number) and the matrix $U^{(s,c)}$ are given in the Appendix.B Clearly, the repulsive Hubbard-interactions enhance(suppress) spin(charge) susceptibility. Note that the RPA approach only works for weak $U^{(s)}$ when all the eigenvalues of the denominator matrix $I-\chi^{(0)}\left(\mathbf{q},i\nu\right)U^{(s)}$ are positive. At the critical interaction strength, the lowest eigenvalue of that matrix touches zero, which leads to divergence of the spin susceptibility and invalidates the RPA, suggesting the formation of magnetic order. The critical strength $U_c$ as function of $J_H/U$ is given in the phase-diagram in Fig.\ref{phase_diagram} as the phase boundary separating the superconducting phase and the magnetic phase. For $U>U_c$, intra-sublattice FM phase emerges since the bare susceptibility shown in Fig.\ref{DOS_sus}(b) peaks at the $\Gamma$-point (the renormalized susceptibility also peaks there), consistent with the ``IOP" phase obtained in Ref.\onlinecite{Wu1}.
\begin{table}
\centering
  \caption{The ten possible pairing symmetries for K$_2$Cr$_3$As$_3$, among which five are spin-singlet while the left are spin-triplet.}
\label{Tab:one}
\begin{tabular}{@{}ccccccccccc@{}}
\\\hline\hline
 singlet   &  triplet   \\
 \hline\hline
 $s$          &  $p_z$  \\
 $(d_{x^2-y^2},d_{xy})$             &      $(d_{x^2-y^2},d_{xy})\cdot p_z$             \\
 $(p_x,p_y)\cdot p_z$         &  $(p_x,p_y)$          \\
 $f_{x^3-3xy^2}\cdot p_z$         &  $f_{x^3-3xy^2}$             \\
 $f_{y^3-3x^2y}\cdot p_z$         &  $f_{y^3-3x^2y}$             \\
 \hline\hline
\end{tabular}
\end{table}

\subsection {Pairing symmetry study}
 When the interaction strength $U<U_c$, there can be short-ranged spin or charge fluctuations in the system. Let's consider a
 Cooper pair with momentum/orbital $(\mathbf{k'}t,-\mathbf{k'}s)$, which could be scattered to $(\mathbf{k}p,-\mathbf{k}q)$ via exchanging spin or charge fluctuations. This second-order perturbation process will contribute an effective interaction vertex $\Gamma^{pq}_{st}(\mathbf{k,k'})$. Considering only intra-band pairings, we project the effective interaction $\Gamma^{pq}_{st}(\mathbf{k,k'})$ onto the FSs and obtain an effective pairing potential $V^{\alpha\beta}(\mathbf{k,k'})$, where $\alpha/\beta=1,\cdots,3$ are band indices. The explicit formula of $\Gamma^{pq}_{st}(\mathbf{k,k'})$ and $V^{\alpha\beta}(\mathbf{k,k'})$ are given in the Appendix.B.

 Finally, one should solve the following linearized gap equation to find the pairing symmetry,
   \begin{equation}
 -\frac{1}{(2\pi)^3}\sum_{\beta}\iint_{FS}
d^{2}\mathbf{k'}_{\Vert}\frac{V^{\alpha\beta}(\mathbf{k,k'})}{v^{\beta}_{F}(\mathbf{k'})}\Delta_{\beta}(\mathbf{k'})=\lambda
 \Delta_{\alpha}(\mathbf{k}).\label{pair_egv}
\end{equation}
Here, the integration and summation are along various FS patches. The $v^{\beta}_{F}(\mathbf{k'})$
is Fermi velocity and $\mathbf{k}'_{\Vert}$ represents the component within patch $\beta$. This equation can be looked upon as an eigenvalue problem, where the eigenvector $\Delta_{\alpha}(\mathbf{k})$ represents the relative gap function near $T_c$, and the eigenvalue $\lambda$ determines $T_c$ via $T_{c}$=cut off energy $e^{-1/\lambda}$. The leading pairing symmetry is determined by the largest eigenvalue $\lambda$. Generally, the RPA study on pairing symmetry only works well for $U<<U_c$, which will be our focus in the following.

The eigenvector(s) $\Delta_{\alpha}(\mathbf{k})$ for each eigenvalue $\lambda$ introduced above form an irreducible representation of the $D_{3h}$ point group of K$_2$Cr$_3$As$_3$. Altogether, there are 10 different pairing symmetries, including 5 singlet pairings and 5 triplet ones, which are listed in Table.\ref{Tab:one}. Here, the $(d_{x^2-y^2},d_{xy})$, $(d_{x^2-y^2},d_{xy})\cdot p_z$, $(p_x,p_y)$ and $(p_x,p_y)\cdot p_z$ symmetries each form a 2D representation of the point group with degenerate pairing eigenvalues. Generally, below $T_c$, the two degenerate eigenvectors of these representations would be mixed as  $d_{x^2-y^2}\pm i d_{xy}$,$(d_{x^2-y^2}\pm i d_{xy})\cdot p_z$, $p_x\pm ip_y$ and $(p_x\pm ip_y)\cdot p_z$. There are two different $f$-wave symmetries, i.e. $f_{x^3-3xy^2}$ and $f_{y^3-3x^2y}$. While the gap functions of both $f-$ waves change signs with every 60$^o$ degree rotation, their nodal lines are different by 30$^o$ rotation. Due to the absence of spin orbital coupling (SOC) in our model, each of the five triplet pairings possesses three degenerate components, i.e. $\uparrow\uparrow, \downarrow\downarrow, (\uparrow\downarrow+\downarrow\uparrow)$, which we shall not distinguish in the following unless otherwise stated. Note that the concrete formalism of the gap function of each symmetry is not given a prior to the RPA calculation. Instead, they should be solved from the gap equation (\ref{pair_egv}).

\begin{figure}
\scalebox{0.32}{{\includegraphics[scale=0.5]{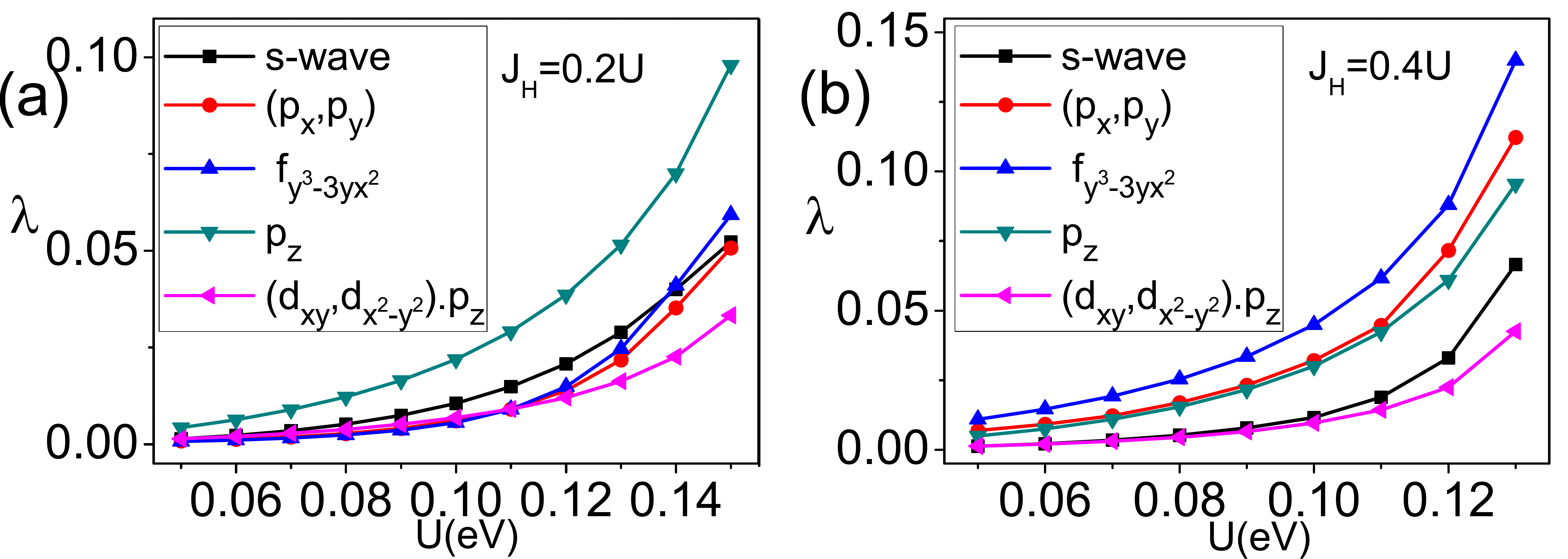}}
}\caption{(color online). The $U$-dependence of the largest eigen values $\lambda$ for different pairing symmetries for (a).$J_H=0.2U$ and (b). $J_H=0.4U$. Only the five stronger symmetries are shown.} \label{U_dependence}
\end{figure}

\begin{figure}
\scalebox{0.32}{{\includegraphics[scale=0.5]{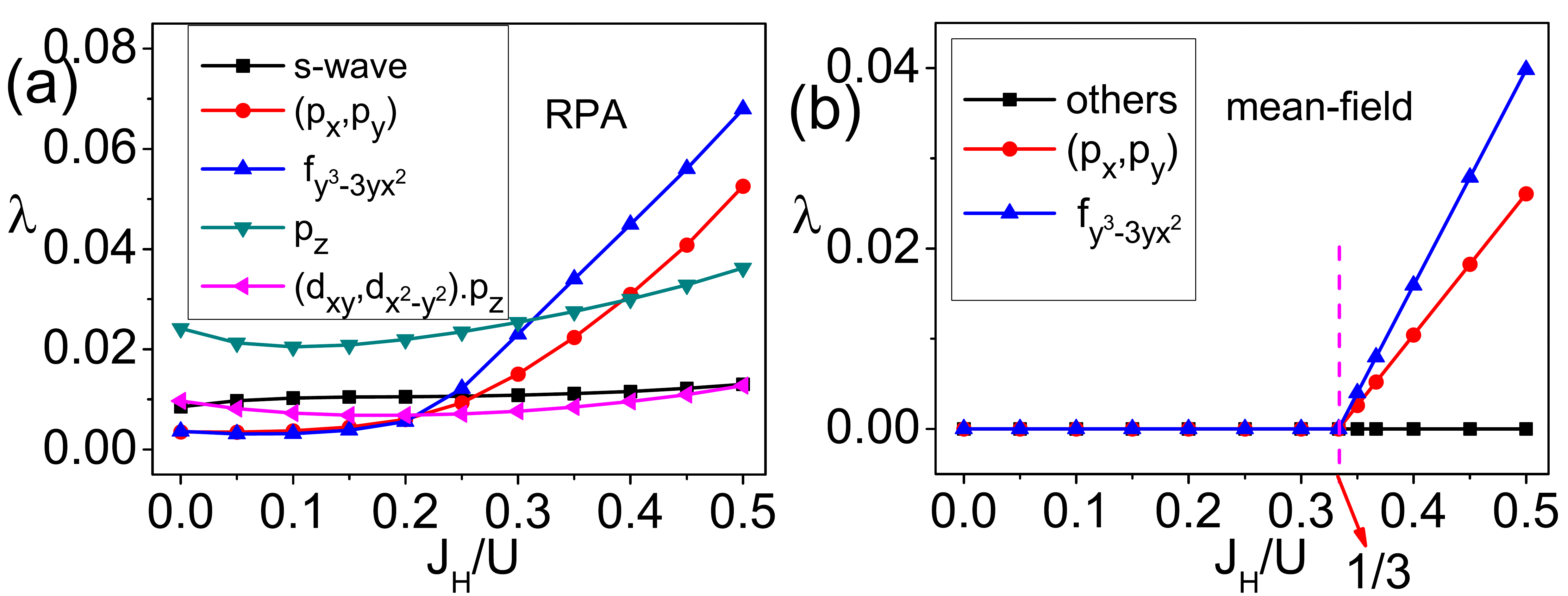}}
}\caption{(color online). The $\frac{J_H}{U}$-dependence of the largest eigen values $\lambda$ for the five stronger pairing symmetries for fixed $U=0.1{\rm eV}$ for (a). RPA and (b). mean-field results.} \label{J_dependence}
\end{figure}

Our RPA calculations were performed on $18\times 18\times 200$ and $30\times 30\times 300$ lattices with periodic boundary condition. The consistence between the two results indicates our RPA result converges in the thermodynamic limit. The $U$-dependence of the largest eigenvalue $\lambda$ for the five stronger pairing symmetries is shown in Fig.\ref{U_dependence} for (a). $J_H=0.2U$ and (b). $J_H=0.4U$. Clearly, all these eigenvalues increase promptly with the enhancement of $U$ and would diverge for $U\to U_c$. For $J_H=0.2U$ shown in (a), the $p_z$-wave is the leading pairing symmetry and dominates other symmetries; for $J_H=0.4U$ shown in (b), the $f_{y^3-3x^2y}$-wave becomes the leading one, with the $(p_x,p_y)$- and $p_z$- waves to be close candidates. Fig.\ref{J_dependence}(a) shows the $\frac{J_H}{U}$-dependence of these eigenvalues for fixed $U=0.1{\rm eV}$. Obviously, the spin-triplet $p_z$, $f_{y^3-3x^2y}$ and $(p_x,p_y)$ symmetries dominate other symmetries in the whole range of $r\equiv \frac{J_H}{U}$. While for $r<r_c$ ($\approx \frac{1}{3}$) the $p_z$-wave symmetry dominates other symmetries, for $r>r_c$ the $f_{y^3-3x^2y}$-wave symmetry becomes the leading symmetry with the $(p_x,p_y)$ and $p_z$-wave states to be close candidates.

\subsection {Triplet pairings}
\begin{figure}
\scalebox{0.3}{\includegraphics[scale=1.0]{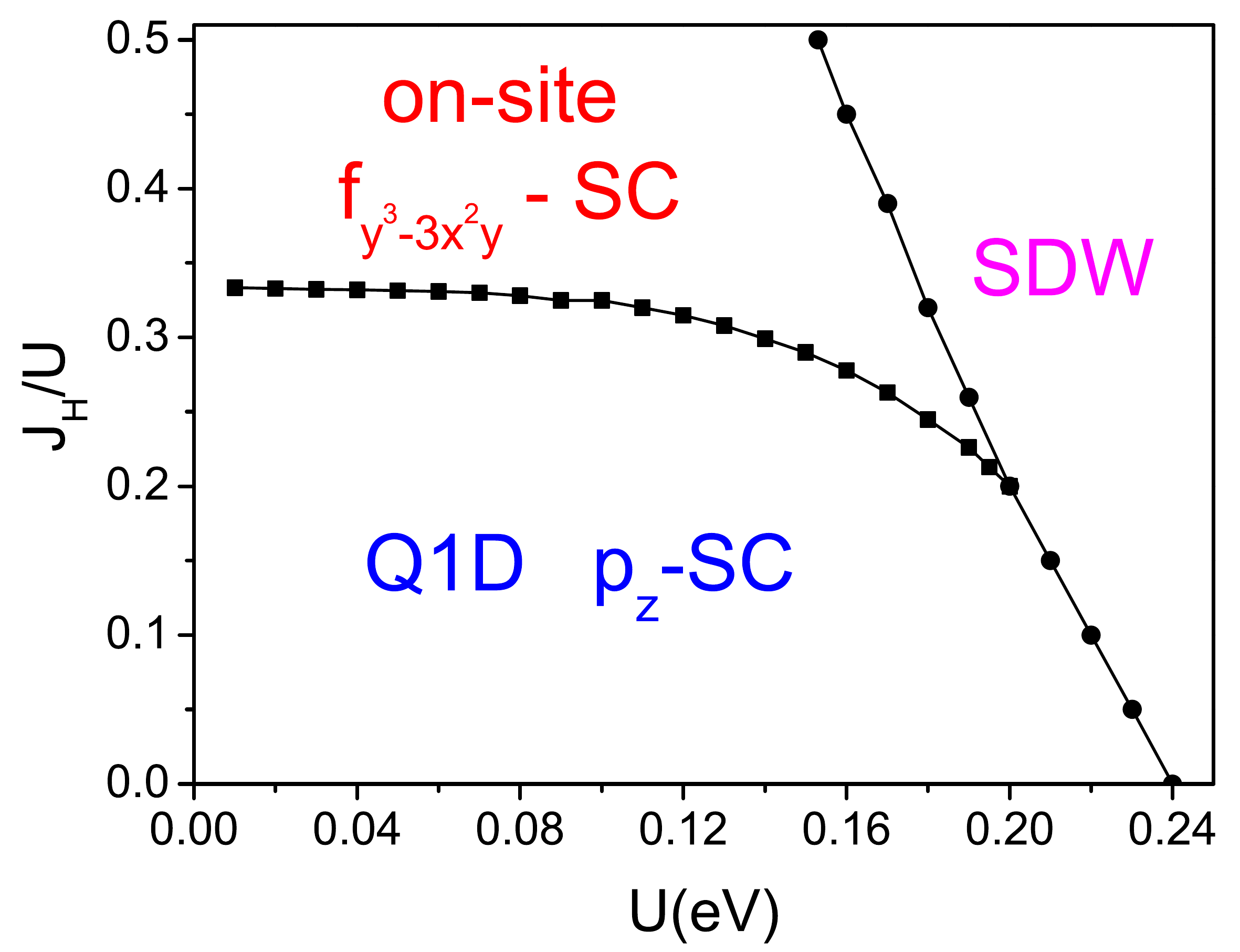}
}\caption{(color online). The ground state phase-diagram in the $U-\frac{J_H}{U}$ plane.} \label{phase_diagram}
\end{figure}

It's interesting to find from the above shown RPA results that all the dominating pairing symmetries in different regimes, i.e. the $p_z$, $f_{y^3-3x^2y}$ and $(p_x,p_y)$ symmetries are triplet pairings. The ground state phase-diagram in the $U-\frac{J_H}{U}$ plane is shown in Fig.\ref{phase_diagram}, where three possible phases are present. For $U>U_c$ (which is around 0.16-0.24 eV and is $J_H/U$-dependent), the intra-sublattice FM SDW order emerges. For $U<U_c$, two triplet pairings, i.e. the $p_z$ and $f_{y^3-3x^2y}$-waves emerge, which are separated by the critical value $r_c$ (which is $U$-dependent) for $r\equiv \frac{J_H}{U}$. For $r$ below or above $r_c$, the $p_z$- or $f_{y^3-3x^2y}$-wave pairing is the leading pairing symmetry respectively. In the limit of $U\to 0$, we have $r_c\to \frac{1}{3}$; with the enhancement of $U$, $r_c$ first decreases slightly and then drops near the SDW critical point. In our focused regime of $U<<U_c$ in RPA, we have $r_c\approx \frac{1}{3}$. While both gap functions have node lines, they are along $k_y=\pm \sqrt{3}k_x, 0$ directions for the $f_{y^3-3x^2y}$-wave pairing and in the $k_z=0$ plane for the $p_z$-wave pairing.
\begin{figure*}
\scalebox{0.4}{{\includegraphics[scale=0.55]{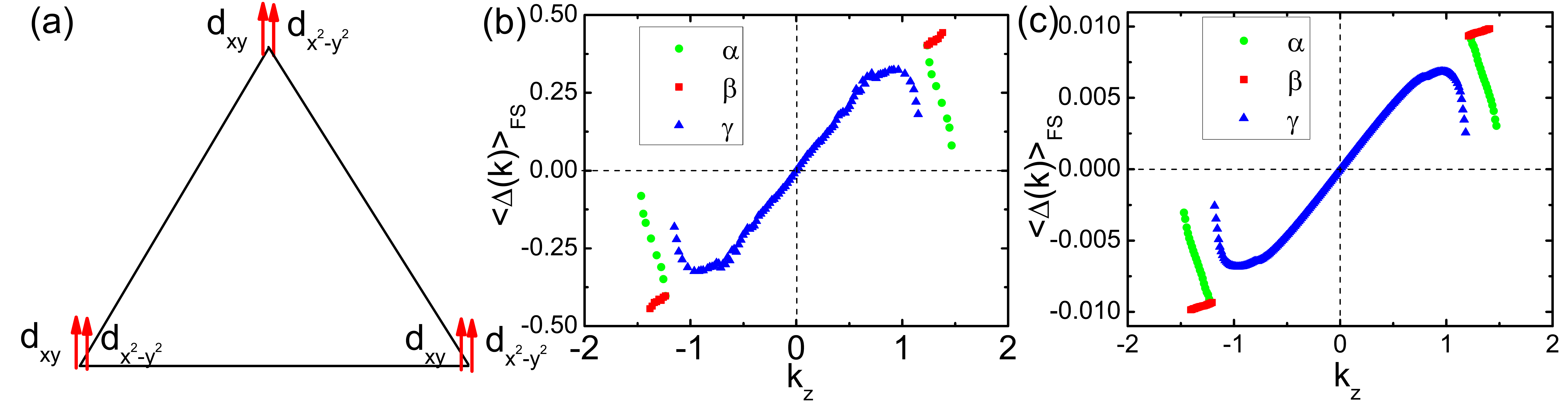}}
}\caption{(color online). (a). Dominating pairing pattern in the real space for the $f_{y^3-3x^2y}$-wave pairing shown in Fig.(\ref{phase_diagram}). Note that only the $\uparrow\uparrow$ pairing component is shown. The $k_z$-dependence of the relative gap function of $p_z$-wave pairing averaged on the FSs for (b). RPA result for the Hubbard-Hund model with $J_{H}/U=0.1$ and (c) Mean-field result for the $t-J$ model.} \label{gap_func}
\end{figure*}

Note that the dominating part of the leading $f_{y^3-3x^2y}$- and sub-leading $(p_x,p_y)$-wave pairings in the regime $r>r_c\approx\frac{1}{3}$ are on-site inter-orbital pairings driven by the Hund's rule coupling. To clarify this point, let's turn off the second-order perturbation process of RPA, and study the problem in the mean-field level. The Hubbard-Hund's interaction (\ref{H-H-model}) can be decoupled in the real space as the sum of an interorbital pairing part
\begin{eqnarray}
H_{\rm inter}&=&\left(V+J_H\right)\sum_{i,\mu<\nu}\hat{\Delta}^
{\left(-\right)\dagger}_{\mu\nu\uparrow\downarrow}\left(i\right)\hat{\Delta}^
{\left(-\right)}_{\mu\nu\uparrow\downarrow}\left(i\right)\nonumber\\&+&\left(V-J_H\right)%\cdot
\Big[\sum_{i,\mu<\nu,\sigma}\hat{\Delta}^
{\dagger}_{\mu\nu\sigma\sigma}\left(i\right)\hat{\Delta}_{\mu\nu\sigma\sigma}\left(i\right)\nonumber \\&+&
\sum_{i,\mu<\nu}\hat{\Delta}^
{\left(+\right)\dagger}_{\mu\nu\uparrow\downarrow}\left(i\right)\hat{\Delta}^
{\left(+\right)}_{\mu\nu\uparrow\downarrow}\left(i\right)\Big] \label{decouple1}
\end{eqnarray}
and an intraorbital pairing part\cite{BiH}
\begin{eqnarray}
H_{\rm intra}&=&U\sum_{i\mu}\hat{\Delta}^{\dagger}_{\mu\mu\uparrow\downarrow}\left(i\right)\hat{\Delta}_{\mu\mu\uparrow\downarrow}\left(i\right)
\nonumber\\&+&J_H\sum_{i,\mu\ne\nu}\hat{\Delta}^{\dagger}_{\mu\mu\uparrow\downarrow}\left(i\right) \hat{\Delta}_{\nu\nu\uparrow\downarrow}\left(i\right),\label{decouple2}
\end{eqnarray}
where $\hat{\Delta}_{\mu\nu\sigma\sigma^{\prime}}(i)\equiv c_{i\mu\sigma}c_{i\nu\sigma^{\prime}}$ and $\hat{\Delta}^{\left(\pm\right)}_{\mu\nu\uparrow\downarrow}\equiv \frac{1}{\sqrt{2}}\left(\hat{\Delta}_{\mu\nu\uparrow\downarrow}\pm\hat{\Delta}_{\mu\nu\downarrow\uparrow}\right)$. Clearly, for repulsive interactions, energetically favored mean-field pairings are only possible in the regime $J_H>V$ (or equivalently $r\equiv\frac{J_H}{U}>r_c=\frac{1}{3}$ upon $U=V+2J_H$). The possible pairings channels include two on-site inter-orbital triplet pairings, which are the leading $f_{y^3-3x^2y}$-wave: $\hat{\Delta}_{23,(\sigma\sigma,\uparrow\downarrow+\downarrow\uparrow)}$ and the subleading $(p_x,p_y)$-doublets: $\left(\hat{\Delta}_{12,(\sigma\sigma,\uparrow\downarrow+\downarrow\uparrow)}, \hat{\Delta}_{13,(\sigma\sigma,\uparrow\downarrow+\downarrow\uparrow)}\right)$, whose pairing eigenvalues as functions of $\frac{J_H}{U}$ are shown in Fig.\ref{J_dependence}(b). The real-space pairing pattern of the mean-field $f_{y^3-3x^2y}$-wave pairing is shown in Fig.\ref{gap_func}(a) (only the $\uparrow\uparrow$ pairing component is shown), where the pairing is between the $d_{xy}$ and $d_{x^2-y^2}$ orbitals. Comparing Fig.\ref{J_dependence}(a) and (b), the FM spin fluctuations considered in RPA further enhance the two triplet pairings and slightly reduce $r_c$, besides extending the pairing coherent length.

\begin{figure*}[tb]
\centerline{\includegraphics[height=7cm]{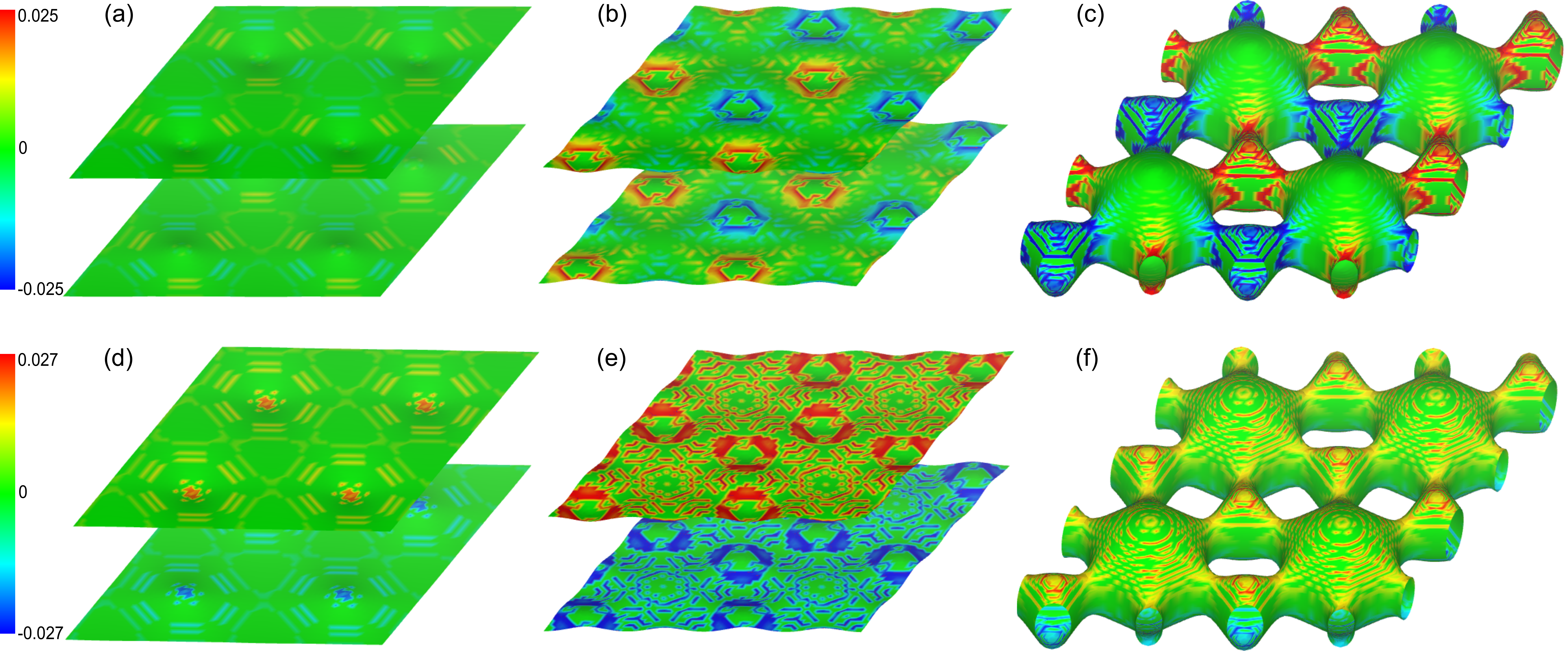}}
\caption{(color online). The relative gap function distribution over the three FSs for the $f_{y^3-3x^2y}$ wave state(top panel) and  the $p_z$ wave state(bottom panel) in RPA calculations. Fig. (a) (d), (b) (e) and (c) (f) are for $\alpha$, $\beta$ and $\gamma$ FSs. With $U=0.1$ eV, the adopted Hund couplings are $J_H=0.4U$ and $J_H=0.1 U$ for $f_{y^3-3x^2y}$ and $p_z$ wave states in the calculations. The gap functions are plotted on the 30$\times$30$\times$300 k-mesh. Note that the strong local oscillations of the gap are artifact of plotting due to finite lattice.   \label{gap3drpa} }
\end{figure*}

 The distribution of the relative gap function of the leading $f$-wave pairing obtained by RPA for $U=0.1{\rm eV}, J_H=0.4U$ is shown on the FSs in Fig.\ref{gap3drpa}(a),(b) and (c). This orbital-singlet spin-triplet pairing state is mirror-reflection even about the $k_z=0$ plane, and its gap function changes sign over every 60$^{o}$ degree rotation about the $z$-axis, causing line gap nodes in the direction of $k_y=0,\pm\sqrt{3}k_x$. While the pairing amplitudes on the $\beta$- and $\gamma$-FSs are comparable, that on the $\alpha$-FS is much weaker as the dominating orbital component on that FS is $d_{z^2}$, which nearly doesn't take part in the pairing, as suggested by the above mean-field analysis.

In real material, the situation $\frac{J_H}{U}>r_c\approx\frac{1}{3}$ can hardly be realized because the orbitals in our effective models are the molecule orbitals that are equally distributed on the three Cr2 atoms and thus the effective value of $J_H$ should be small. Therefore, the realized pairing symmetry in K$_2$Cr$_3$As$_3$ might more probably be the $p_z$-wave pairing displayed in the regime $r<r_c$ in the phase-diagram Fig.\ref{phase_diagram}, which is consistent with the mean-field study below on the $t-J$ model of the system. Since the Hund's rule coupling driven pairings don't include the $p_z$-wave, as suggested by Fig.\ref{J_dependence}(b), this pairing is purely driven by the intra-sublattice FM spin fluctuations reflected in Fig.\ref{DOS_sus}(b).

 The distribution of the relative gap function of the leading $p_z$-wave pairing obtained by RPA for $U=0.1{\rm eV}, J_H=0.1U$ is shown on the FSs in Fig.\ref{gap3drpa}(d), (e) and (f). This triplet pairing is mirror-reflection odd about the $k_z=0$ plane, causing line gap nodes on that plane on the 3D $\gamma$-FS. The gap function of this pairing is isotropic on the FS sheets with a fixed $k_z$ and its $k_x$- and $k_y$-dependence is weak; particularly it doesn't change sign all over the regime $k_z>0$ (or $k_z<0$) on the FSs, causing no extra gap nodes. Obviously, the gap amplitude on the Q1D $\beta$-FS dominates those of $\alpha$- and $\gamma$-FSs, suggesting the Q1D nature of the obtained $p_z$-wave pairing. Considering the orbital components of $\beta$-FS, this pairing takes place mainly within the $d_{xy}$ and $d_{x^2-y^2}$ orbitals. To exhibit the $k_z$-dependence of the obtained $\Delta_{\alpha}(\mathbf{k})$, we plot in Fig.\ref{gap_func}(b) $\langle\Delta_{\alpha}(\mathbf{k})\rangle_{FS}$, i.e. the averaged gap function on the FS as function of the fixed $k_z$, in comparison with that obtained in the following strong-coupling limit in Fig.\ref{gap_func}(c). The consistency between the results obtained from the two approaches are quite good.
\section{The strong coupling limit}

In the strong coupling limit, we consider super-exchange type of interactions. From DFT calculations on K$_2$Cr$_3$As$_3$\cite{Wu1}, the NN and NNN exchange couplings along the Cr-chain are antiferromagnetic (AFM) and FM respectively. As only the Cr2 atoms are included in our three-band model, the super-exchange interactions between adjacent Cr2 along the chain direction (i.e. the $z$-direction) are actually FM, which provides us the following exchange interaction term
\begin{eqnarray}
H_{\rm J}=-J_2\sum_{\langle ij \rangle, \alpha}\textbf{S}_{\alpha i}\cdot\textbf{S}_{\alpha j},\label{exchange}
\end{eqnarray}
with $J_2>0$. Here $\alpha \in \{d_{z^2},d_{xy},d_{x^2-y^2}\}$, $\langle ij \rangle$ denotes NN bond along the $z$-direction in our model but NNN bond in real materials and $\textbf{S}_{\alpha i}=\frac{1}{2}\sum_{\sigma\sigma'}c^\dag_{\alpha i\sigma}\bm{\sigma}_{\sigma\sigma'} c_{\alpha i\sigma'}$ is the local spin operator. The inter-chain exchange coupling has been neglected here due to the Q1D character of the material. Note that only intra-orbital  FM exchange coupling is considered because of the ``IOP" spin configuration introduced in Ref.\onlinecite{Wu1}. Adding this interaction term into the tight-binding model, we obtain an effective $t-J$ model. Since the band filling here is far from making the system a Mott-insulator even in the large U limit, we shall omit the no-double-occupance constraint and perform a mean-field analysis on the model.

%For the strong-coupling limit,
We consider the super-exchange interaction Eq.(\ref{exchange}). The Hamiltonian of our model is,
\begin{eqnarray}
H=H_{\rm TB}+H_{\rm J}.\label{t-J}
\end{eqnarray}
 Since this material is far from Mott-insulator, we neglect the no-double-occupance constraint on this $t-J$ model and perform a mean-field study on it.

In terms of real-space pairing operators, this interaction Hamiltonian can be written as,
\begin{eqnarray}
H_{\rm J}&=&\frac{3}{4}J_{2}\sum_{\langle ij \rangle,\mu}P^{\dag}_{ij \mu}P_{ij \mu}-\frac{1}{4}J_{2}\sum_{\langle ij \rangle,\mu\gamma}T^\dag_{ij \mu,\gamma}T_{ij \mu,\gamma},\nonumber\\
P^{\dagger}_{ij \mu}&=&\frac{1}{\sqrt{2}}(c^{\dag}_{i \mu\uparrow}c^\dag_{j \mu\downarrow}-c^\dag_{i \mu\downarrow}c^\dag_{j \mu\uparrow}),\nonumber\\
T^{\dagger}_{ij \mu,\gamma}&=&
\begin{cases}
c^\dag_{i \mu\uparrow}c^\dag_{j \mu\uparrow}   & \gamma=1,\\
\frac{1}{\sqrt{2}}(c^\dag_{i \mu\uparrow}c^\dag_{j \mu\downarrow}+c^\dag_{i \mu\downarrow}c^\dag_{j \mu\uparrow}) & \gamma=0, \\
c^\dag_{i \mu\downarrow}c^\dag_{j \mu\downarrow} &\gamma=-1. \label{decouple}
\end{cases}
%T_{0,ij}&=&\frac{1}{\sqrt{2}}(c^\dag_{i\uparrow}c_{j\downarrow}+c^\dag_{i\downarrow}c_{j\uparrow})\\
%T_{1,ij}&=&c^\dag_{i\uparrow}c_{j\uparrow}\\
%T_{-1,ij}&=&c^\dag_{i\downarrow}c_{j\downarrow}
\end{eqnarray}
Here $P_{ij \mu}$ is the spin-singlet pairing operator and $T_{ij \mu,\gamma}$ is the spin-triplet one with $S_z=\gamma$. From Eq. (\ref{decouple}), it's clear that for $J_2>0$ ($J_2<0$), spin-triplet (singlet) pairing will be favorable. In K$_2$Cr$_3$As$_3$, FM intra-sublattice spin correlation leads to $J_2>0$, thus we only consider the triplet pairing terms. Due to the absence of SOC in our model, the pairing states in $S_z=-1,0,1$ channels are degenerate, which allows us to only consider the $S_z=0$ channel in the following.

Defining $\Delta_{\mu}\equiv -iJ_2\langle T_{i,i+\hat{z},\mu,0}\rangle/2\sqrt{2}$ as the (real) pairing order parameters, the above $H_J$ term can be decoupled as
\begin{equation}
H_{\rm J}=\frac{\sqrt{2}}{2}\sum_{\langle ij\rangle,\mu}\left(-i\Delta_{\mu}T^{\dagger}_{ij\mu,0}+h.c\right)+\frac{2N}{J_2}\sum_{\mu}|\Delta_{\mu}|^2.\label{decoupling}
\end{equation}
In combination with the tight-binding part, the total mean-field Hamiltonian in the momentum space is
\begin{eqnarray}
H_{\rm mf}&=&\sum_{\mathbf{k}}\Psi^\dag(\mathbf{k})H(\mathbf{k})\Psi(\mathbf{k})+\frac{2N}{J_2}\sum_{\mu}|\Delta_{\mu}|^2,\nonumber\\
H(\mathbf{k})&=&\left(\begin{array}{cc}
h(\mathbf{k}) & \Delta_{\uparrow\downarrow}(\mathbf{k}) \\
 \Delta^{\dagger}_{\uparrow\downarrow}(\mathbf{k}) & -h^{*}(-\mathbf{k}) \\
 \end{array}\right), \nonumber\\
 \Delta_{\uparrow\downarrow}(\mathbf{k})&=&\left(\begin{array}{ccc}
\Delta_{1}(\mathbf{k}) &  &  \\
  & \Delta_{2}(\mathbf{k}) & \\
  &   &  \Delta_{3}(\mathbf{k})  \\
 \end{array}\right),
\end{eqnarray}
where $\Psi^\dag(\mathbf{k})=[c^{\dag}_{1\uparrow}(\mathbf{k}),c^{\dag}_{2\uparrow}(\mathbf{k}),c^{\dag}_{3\uparrow}(\mathbf{k}),c_{1\downarrow}(-\mathbf{k}),c_{2\downarrow}(-\mathbf{k}),\\c_{3\downarrow}(-\mathbf{k})]$ and $\Delta_{\mu}(\mathbf{k})=\Delta_{\mu}{\rm sin}k_z$. Note that a constant $\sum_{\mathbf{k}\mu}[h_{\mu\mu}(\mathbf{k})-\mu_c]$, which has no contribution to the dynamics, has been neglected on the above.

\begin{figure}[tb]
\centerline{\includegraphics[height=4.2 cm]{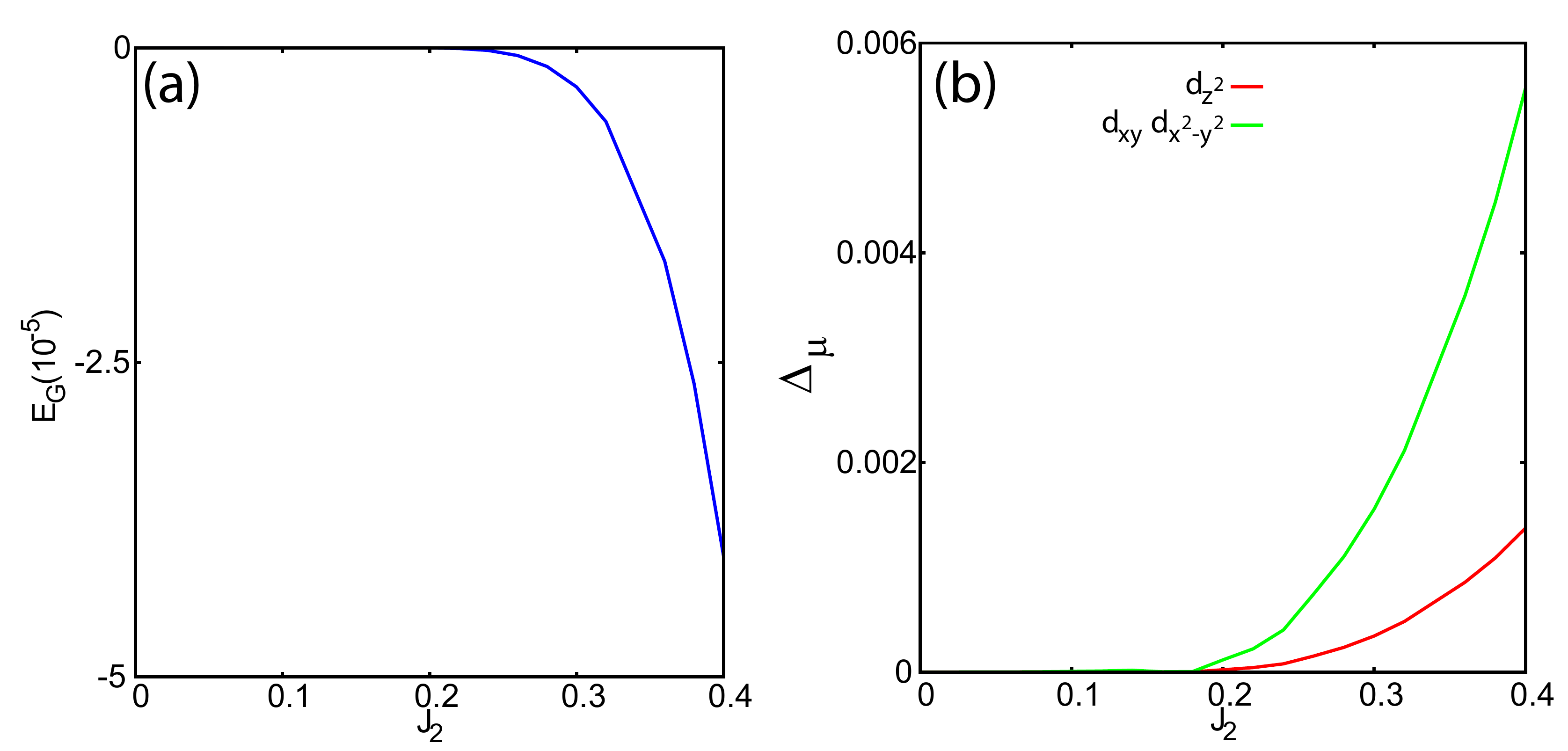}}
\caption{(color online). The ground state energy $E_G$ (a) and gap amplitudes $\Delta_{\mu}$ (b) for triplet pairing in $S_z=0$ channel as a function of $J_2$.   \label{sc} }
\end{figure}

\begin{figure*}[tb]
\centerline{\includegraphics[height=3.3cm]{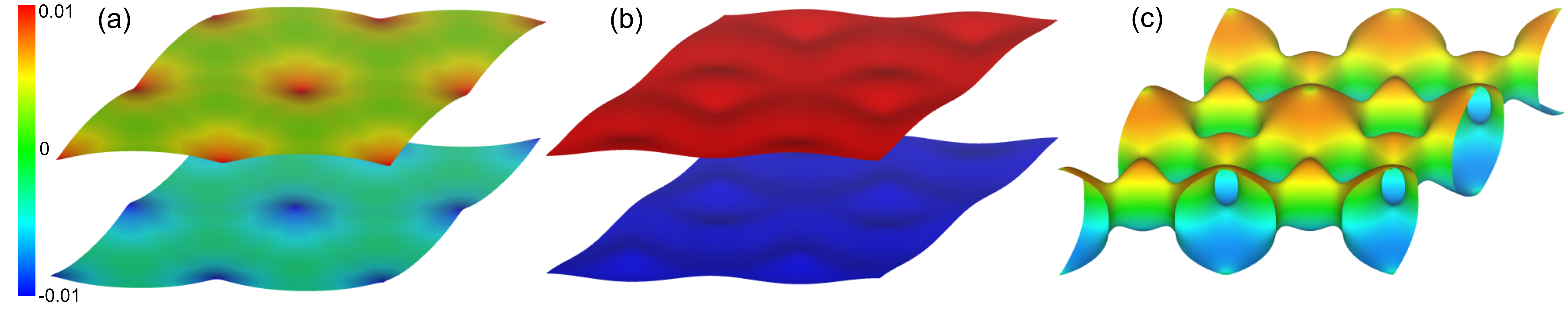}}
\caption{(color online). The gap function distribution over the three FSs  with $\Delta^{\uparrow\downarrow}_{z^2}=0.0025 {\rm eV}, \Delta^{\uparrow\downarrow}_{x^2-y^2}=\Delta^{\uparrow\downarrow}_{xy}=0.01 {\rm eV}$, which are twice the gap values obtained for $J_2=0.4{\rm eV}$. Fig. (a), (b) and (c) are for $\alpha$, $\beta$ and $\gamma$ FSs.  \label{gap3d} }
\end{figure*}

\begin{figure}[tb]
\centerline{\includegraphics[height=5.4 cm]{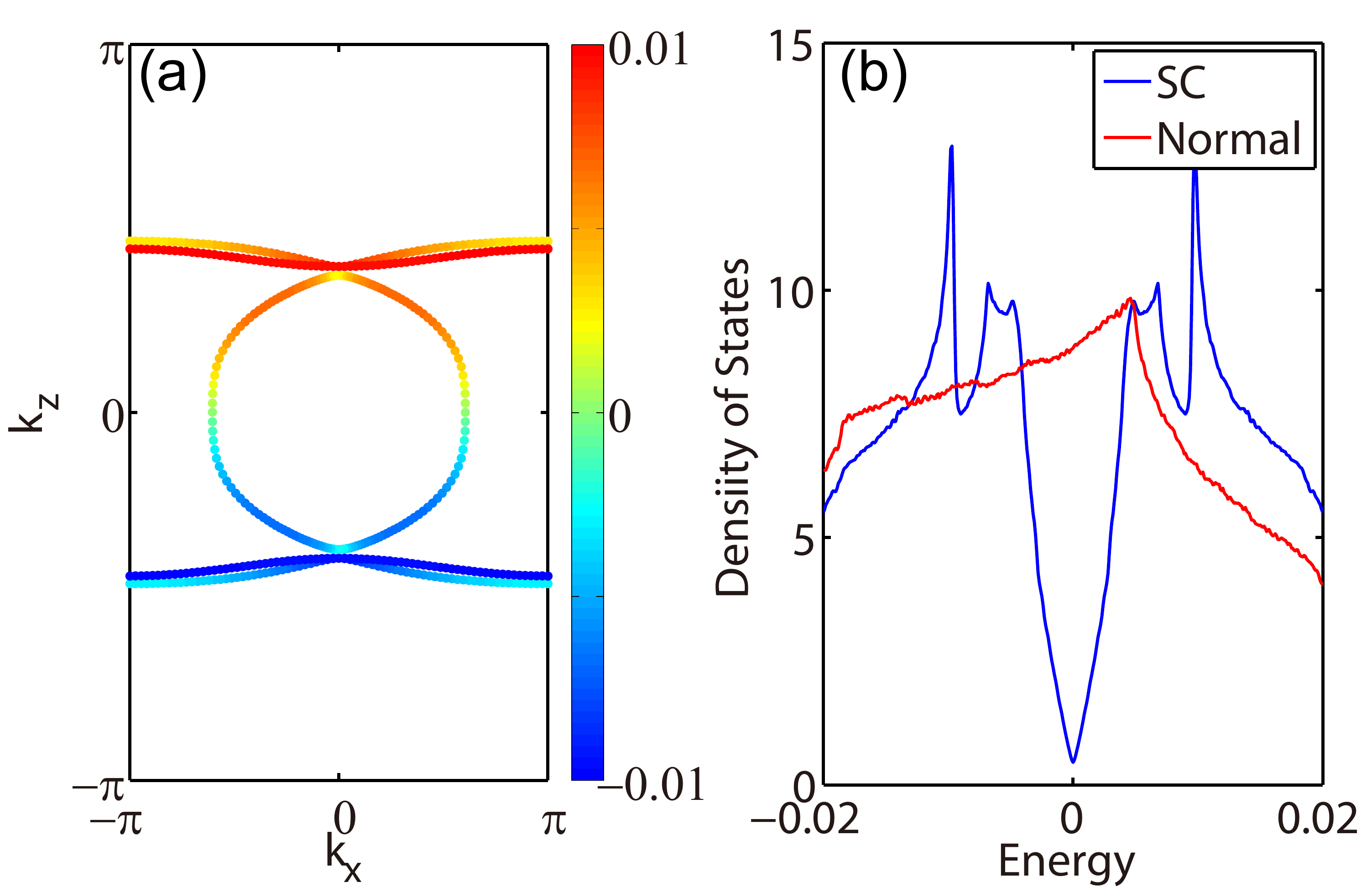}}
\caption{(color online). The SC gap function distribution over the FSs in the $k_y=0$ plane (a) and density of states (for one spin specie) (b) with $\Delta^{\uparrow\downarrow}_{z^2}=0.0025 {\rm eV}, \Delta^{\uparrow\downarrow}_{x^2-y^2}=\Delta^{\uparrow\downarrow}_{xy}=0.01 {\rm eV}$, which are twice the gap values obtained for $J_2=0.4{\rm eV}$. The red and blue lines represent DOS for normal and superconducting states, respectively.   \label{gap1} }
\end{figure}

\begin{figure}
\scalebox{0.32}{\includegraphics[scale=1.0]{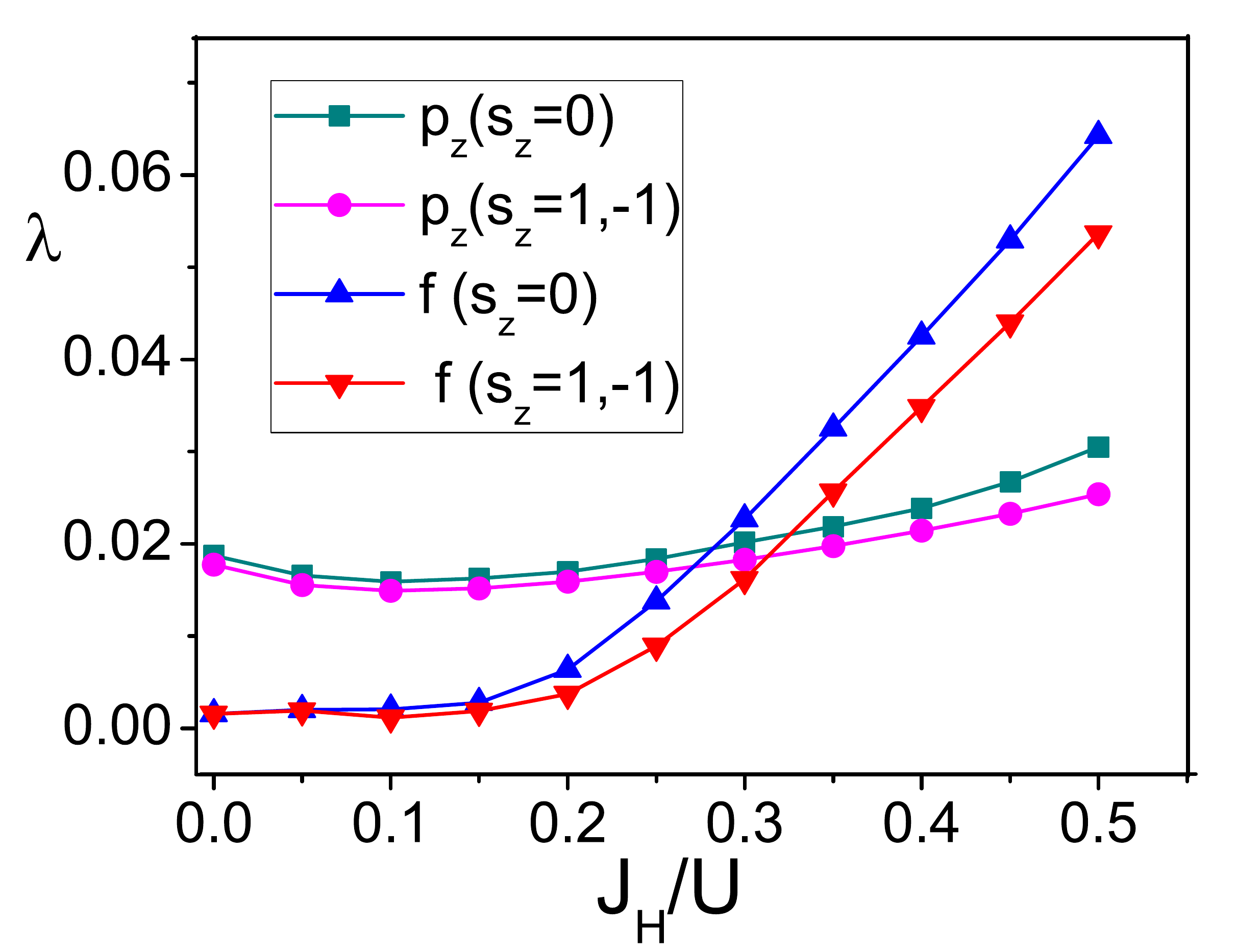}
}\caption{(color online). The $\frac{J_H}{U}$-dependence of the largest eigen values $\lambda$ for the $p_z$- and $f_{y^3-3x^2y}$- wave pairings for fixed $U=0.1$ eV and $\lambda_{so}=10$ meV for the $S_z=0$ ($\uparrow\downarrow$) component and $S_z=\pm 1$ one.} \label{J_dependence_SOC}
\end{figure}

The above mean-field Hamiltonian can be solved via diagonalizing $H(\mathbf{k})$ by a unitary transformation: $U^{\dag}(\mathbf{k})H(\mathbf{k})U(\mathbf{k})={\rm Diag}(E_i)$, with the eigenvalues $E_i=-E_{i+3}>0(i=1,2,3)$. The self-consistent gap equations are,
\begin{equation}
\Delta_{\mu}=\frac{-J_2}{2N}\sum_{\mathbf{k}}\sum_{\alpha=1}^{6}U^{*}_{\mu+3,\alpha}(\mathbf{k})U_{\mu,\alpha}(\mathbf{k})f[E_{\alpha}(\mathbf{k})]{\rm sin}k_z
\end{equation}
where $f(E)$ is the Fermi-Dirac distribution function and we only study the case of zero temperature. Solving these gap equations, we plot the ground state energy $E_G$ and the three gap amplitudes $\Delta_{\mu}$ ($\mu=1,\cdots,3$) as function of $J_2$ in Fig.\ref{sc}(a) and (b) respectively. Clearly, the gap amplitudes of $d_{xy}$ and $d_{x^2-y^2}$ orbitals, belonging to the $E'$ irreducible representation, are equal and dominate that of $d_{z^2}$, which is consistent with the RPA result. Furthermore, there is no phase difference between the SC order parameters of the three orbitals.

Fig.\ref{gap3d} shows the distribution of the pairing gap function $J_2\langle c_{\alpha\downarrow}(-\mathbf{k})c_{\alpha\uparrow}(\mathbf{k}) \rangle=J_2\sum_{\mu}\langle c_{\mu\downarrow}(-\mathbf{k})c_{\mu\uparrow}(\mathbf{k}) \rangle |\xi_{\mu\alpha}(\mathbf{k})|^2=4\sum_{\mu}\Delta_{\mu}|\xi_{\mu\alpha}(\mathbf{k})|^2{\rm sin}(k_z)$ over the FSs in the Brillioun-Zone. Here $\alpha$ is the band index and only intra-band pairing is shown. Clearly, due to the same sign among different $\Delta_{\mu}$, there is no sign change of the gap function all over the regime $k_z>0$ and gap nodes only exist in the $k_z=0$ plane in the 3D $\gamma$-band. Obviously, the pairing symmetry of this superconducting state is $p_z$ since its gap function is reflection odd respect to the mirror plane $k_z=0$ and is rotation symmetric about the $z$-axis. What's more, due to the dominance of $\Delta_{2,3}$ over $\Delta_{1}$, the pairing amplitude on the Q1D $\beta$-band is obviously stronger than that on the $\alpha$-band. Fig.\ref{gap1}(a) shows the gap distribution over the FSs in the $k_x-k_z$ plane with $k_y=0$ and Fig.\ref{gap_func}(c) shows the averaged gap function $\langle\Delta_{\alpha}(\mathbf{k})\rangle_{FS}$ on the FSs with fixed $k_z$ in comparison with the same quantity obtained from RPA in Fig.\ref{gap_func}(b). The obtained gap structure here is well consistent with the results of RPA introduced above. In Fig.\ref{gap1}(b), we plot the DOS in the superconducting state with $\Delta_{1}=2.5{\rm meV}, \Delta_{2}=\Delta_{3}=10{\rm meV}$ obtained for $J_2=0.4{\rm eV}$, scaled by a factor of 2 for clarity. The DOS displays a V-shaped structure, which can be detected by tunneling experiments.

Note that there can be another $p_z$-wave solution of the gap equation with the gap sign of $A'_{1}-d_{z^2}$-orbital different from that of $E' (d_{xy},d_{x^2-y^2})$-orbitals. Due to the phase difference and orbital characters, the gaps on the $\alpha$- and $\gamma$- FSs are strongly anisotropic and line nodes appear on the $\alpha$- FS; but the gap on the $\beta$- FS is quite uniform.
Compared with the former case, there is an additional coherence peak in DOS. This state, however, is energetically unfavorable in our results, due to the extra gap nodes on the FS.\cite{Hu2012}.

\section{Discussion and conclusion}

In conclusion, we have performed combined weak-coupling RPA and strong coupling mean-field study on the pairing symmetry of K$_2$Cr$_3$As$_3$, starting from the three-band tight-binding model. The weak-coupling RPA study on the Hubbard-Hund model reveals two possible triplet pairings in different parameter regimes. For $r\equiv \frac{J_H}{U}>r_c\approx \frac{1}{3}$, on-site inter-orbital pairing (between $d_{xy}$ and $d_{x^2-y^2}$) triplet SC would be driven by the Hund's rule coupling. This pairing has nodal lines in the direction of $k_y=0,\pm \sqrt{3}k_x$.    However, as  the orbitals in our  effective models are the molecule orbitals that are equally distributed on the three Cr2 atoms, the effective value of $J_H$ should be small. Therefore,  $r<r_c$ is a more realistical parameter. For $r<r_c$, the Q1D $p_z$-wave pairing would be driven by the strong intra-sublattice FM spin fluctuations. The $p_z$-wave pairing state obtained by RPA for the Hubbard-Hund model is in well consistency with that obtained by mean-field study on the $t-J$ model in the strong coupling limit. The gap function of this pairing state doesn't change sign in the $k_z>0$ (or $k_z<0$) regime on the FSs and the only line gap nodes lie on the $k_z$=0 plane on the 3D $\gamma$-FS. The pairing amplitude on the Q1D $\beta$- FS dominates those on the Q1D $\alpha$- FS and the 3D $\gamma$- FS, revealing the dominating part of this pairing to be the Q1D pairing within the $E'(d_{xy},d_{x^2-y^2})$ orbitals along the $z$-axis.

In K$_2$Cr$_3$As$_3$, the weak SOC will lift up the degeneracy among the three components of the triplet pairings. The only symmetry allowed on-site SOC term added to our three-band tight-binding model takes the following formalism,
\begin{equation}
 H_{SOC}=i\lambda_{so}\sum_{\mathbf{k}\sigma}\sigma \left[c^{\dag}_{2\sigma}(\mathbf{k})c_{3\sigma}(\mathbf{k})-c^{\dag}_{3\sigma}(\mathbf{k})c_{2\sigma}(\mathbf{k})\right], \label{SOC}
\end{equation}
where a coupling constant $\lambda=10{\rm meV}$ is adopted in our calculations. Classified according to the $S_z$ quantum number of Cooper pairing, possible pairing components include the $S_z=\pm 1$ (i.e. $\uparrow\uparrow, \downarrow\downarrow$) components and the $S_z=0$ (i.e. $\uparrow\downarrow$) component (note that due to the absence of inversion symmetry in K$_2$Cr$_3$As$_3$, the triplet $\uparrow\downarrow+\downarrow\uparrow$ component in principle is allowed to be mixed with the singlet $\uparrow\downarrow-\downarrow\uparrow$ component\cite{Frigeri2004}, although such mixing is weak due to the weak SOC in 3d electronic materials). The $\frac{J_H}{U}$- dependence of the largest pairing eigenvalues $\lambda$ of the $S_z=\pm 1$ and the $S_z=0$ components of the $p_z$- and the $f_{y^3-3x^2y}$- wave pairings obtained from RPA (details of the formula of RPA with SOC can be found in Appendix.B.) is shown in Fig.\ref{J_dependence_SOC}, with fixed $U=0.1{\rm eV}$. Clearly, for both symmetries, the $S_z=0$ component slightly wins over the $S_z=\pm 1$ ones. Our mean-field calculations for the $t-J$ model also energetically prefer the $S_z=0$ component to be the true ground state with a small energy gain.

Note that the gap amplitude on the Q1D $\beta$-band dominates over that on the other two bands in the $p_z$-wave pairing state obtained here, which suggests that Cr2 is more important than Cr1 in determining the pairing properties of the system.  This self-consistently justifies our three-band model as a good start point for the study of pairing symmetry. Nevertheless, to obtain more detailed physical properties of the system, particularly the magnetic property, the six-band model with both Cr1 and Cr2 involved is necessary, as both of them participate in the magnetization and the relative orientation between their magnetic moments are opposite\cite{Wu1}. We leave this problem for future study.

On the experimental aspect, the line gap nodes of this $p_z$-wave pairing manifest themselves in the specific heat\cite{KCrAs,RbCrAs}, the NMR relaxation-rate\cite{NMR_KCrAs}, and the penetration-depth\cite{pen_depth}. Besides, a direct ARPES observation will reveal the existence of line gap nodes on the $k_z=0$ plane on the FS. The feature of triplet pairing in this material is consistent with the strong upper critical field largely beyond the Pauli limit\cite{KCrAs,RbCrAs}. In addition, it will be reflected in the lack of obvious change in the NMR knight-shift upon the superconducting transition\cite{Ishida1998}. Furthermore, the spin triplet pairing state here should be very sensitive to nonmagnetic impurities. The Q1D pairing nature of this $p_z$-wave SC can be detected by anisotropic coherent length. In addition, a direct ARPES observation can reveal the dominance of the gap amplitudes on the Q1D $\beta$-band over the other two bands.

{\bf Acknowledgments}: This work is supported in part by  MOST of China (2012CB821400,2011CBA00100,2015CB921300), NSFC(11190020,91221303,11334012,11274041) and  ``Strategic Priority Research Program (B)" of the Chinese Academy of Sciences( XDB07020200). F.Y is also  supported by the NCET program under Grant No. NCET-12-0038.

{\it Note added}:  After the completion of the present work, we noticed the work by Yi Zhou, {\em et al}\cite{Zhou2015}, studying topics that partly overlap with the present work, but their conclusions are different from ours. After comparing their work with ours, we notice the following difference. Firstly, the band structure adopted in our study is different from theirs, particularly in the aspect of orbital components on each FS. Secondly, in the RPA study, we obtained the relative gap function for each pairing symmetry via solving the linearized gap equation, which guarantees that the leading gap form-factor is not missed. In addition, the issue of critical interaction strength in RPA has been thoroughly investigated in our study, which guarantees that the interaction parameters adopted in our calculation is below the critical one and the RPA is valid.

\newpage
\begin{widetext}
\appendix

\section{The tight-binding model}
From the bandstructure of DFT, we find that there are five bands near the Fermi level and the oribtal characters are mainly $d_{z^2}$, $d_{xy}$ and $d_{x^2-y^2}$. To obtain an effective model, we consider a virtual atom at the center of each Cr triangle with three orbitals, which have the same symmetry as $A'_{1}$ ($d_{z^2}$), and $E'$ ($d_{xy}$, $d_{x^2-y^2}$), as shown in Fig.\ref{model} in the main text. Now the structure of K$_2$Cr$_3$As$_3$ is simplified and every chain has two sublattices. Due to the asymmetry of K atoms and the absence of inversion symmetry, there are two inequivalent sublattices which are labeled as $A$ and $B$ respectively. The tight binding Hamiltonian is
\begin{eqnarray}
h(\mathbf{k})=\left(\begin{array}{cccccc}
h_{11}^{AA}& h_{12}^{AA} & h_{13}^{AA} & h_{11}^{AB} & h_{12}^{AB} &h_{13}^{AB}
 \\
 & h_{22}^{AA} & h_{23}^{AA} & h_{21}^{AB} & h_{22}^{AB} &h_{23}^{AB} \\
 &               & h_{33}^{AA} & h_{31}^{AB} & h_{32}^{AB} &h_{33}^{AB} \\
  &               &              & h_{11}^{BB} & h_{12}^{BB} &h_{13}^{BB} \\
  &               &              &               & h_{22}^{BB} &h_{23}^{BB} \\
  &               &              &               &               &h_{33}^{BB} \\
 \end{array}\right) ,\label{hk}
\end{eqnarray}
where the unshown matrix elements can be obtained from the shown ones by Hermicity of the $h(\mathbf{k})$ matrix. On the above, the orbital indices $1,2$ and $3$ denote $d_{z^2}$, $d_{xy}$ and $d_{x^2-y^2}$ respectively.

 Let $x=\frac{\sqrt{3}}{2}{k_xa_0}$, $y=\frac{1}{2}k_ya_0$ and $z=\frac{1}{2}k_zc_0$, the Hamiltonian matrix elements are as following.

(1). The hoppings between $A'_{1}(d_{z^2})$ orbital and $A'_{1}(d_{z^2})$ orbital lead to the following matrix elements of $h(\mathbf{k})$\\
\begin{eqnarray}
h_{11}^{AA/BB}&=&\epsilon_{1/3}+2\sum_{i=1}^{3}s^{11}_{z,2i}{\rm cos}2iz+(s^{11}_y+2s^{11}_{yz}{\rm cos} 2z)(2{\rm cos}2y+4{\rm cos} x {\rm cos} y )\nonumber\\
h_{11}^{AB}&=&2\sum_{i=1}^{4}s^{11}_{z,2i-1}{\rm cos}[(2i-1)z]+2s^{11}_{yz}{\rm cos} z (2{\rm cos} 2y+4{\rm cos} x {\rm cos} y)\nonumber\\
h_{11}^{BA}&=&h_{11}^{AB,*}=h_{11}^{AB}.
\end{eqnarray}
(2). The hoppings between $A'_{1}(d_{z^2})$ orbital and $E' (d_{xy},d_{x^2-y^2})$ orbitals lead to
\begin{eqnarray}
h_{12}^{AA/BB}&=&2is^{12}_{1y}{\rm sin} 2y +2is^{12}_{1y}{\rm sin} y {\rm cos} x +2\sqrt{3}s^{12}_{2y}{\rm sin} y {\rm sin} x\nonumber\\
h_{21}^{AA/BB}&=&h_{12}^{AA/BB,*}\nonumber\\
h_{13}^{AA/BB}&=&2s^{12}_{2y}{\rm cos} 2y -2i\sqrt{3}s^{12}_{1y}{\rm cos} y {\rm sin} x -2s^{12}_{2y}{\rm cos} y {\rm cos} x\nonumber\\
h_{31}^{AA/BB}&=&h_{13}^{AA/BB,*}\nonumber\\
h_{12}^{AB}&=&2{\rm cos} z (2is^{12}_{1yz}{\rm sin} 2y +2is^{12}_{1yz}{\rm sin} y {\rm cos} x +2\sqrt{3}s^{12}_{2yz}{\rm sin} y {\rm sin} x)\nonumber\\
h_{21}^{AB}&=&h_{12}^{AB,*},
h_{12}^{BA}=h_{12}^{AB},
h_{21}^{BA}=h_{12}^{AB,*}\nonumber\\
h_{13}^{AB}&=&2{\rm cos} z (2s^{12}_{2yz}{\rm cos} 2y -2\sqrt{3}is^{12}_{1yz}{\rm cos} y {\rm sin} x -2s^{12}_{2yz}{\rm cos} y {\rm cos} x)\nonumber\\
h_{31}^{AB}&=&h_{13}^{AB,*},
h_{13}^{BA}=h_{13}^{AB},
h_{31}^{BA}=h_{13}^{AB,*}
\end{eqnarray}
(3). The hoppings between $E' (d_{xy},d_{x^2-y^2})$ orbitals and $E' (d_{xy},d_{x^2-y^2})$ orbitals lead to
\begin{eqnarray}
h_{22}^{AA/BB}&=&\epsilon_{2/4}+2s^{22}_{11y}{\rm cos} 2y+(s^{22}_{11y}+3s^{22}_{22y}){\rm cos} x {\rm cos} y +2\sum_{i=1}^{2}s^{22}_{z,2i}{\rm cos}2iz\nonumber\\
h_{23}^{AA/BB}&=&\sqrt{3}(s^{22}_{11y}-s^{22}_{22y}){\rm sin} x {\rm sin} y +2is^{22}_{12y}{\rm sin} 2y -4is^{22}_{12y}{\rm cos} x {\rm sin}y\nonumber\\
h_{32}^{AA/BB}&=&h_{23}^{AA/BB,*}\nonumber\\
h_{33}^{AA/BB}&=&\epsilon_{2/4}+2s^{22}_{22y}{\rm cos} 2y+(3s^{22}_{11y}+s^{22}_{22y}){\rm cos} x {\rm cos} y +2\sum_{i=1}^{2}s^{22}_{z,2i}{\rm cos}2iz\nonumber\\
h_{22}^{AB}&=&2{\rm cos} z\left[2s^{22}_{11yz}{\rm cos} 2y+(s^{22}_{11yz}+3s^{22}_{22yz}){\rm cos} x {\rm cos} y\right]+2\sum_{i=1}^{3}s^{22}_{z,2i-1}{\rm cos}(2i-1)z \nonumber\\
h_{22}^{BA}&=&h_{22}^{AB,*}=h_{22}^{AB}\nonumber\\
h_{23}^{AB}&=&2{\rm cos} z\left[\sqrt{3}(s^{22}_{11yz}-s^{22}_{22yz}){\rm sin} x {\rm sin} y +2is^{22}_{12yz}{\rm sin} 2y -4is^{22}_{12yz}{\rm cos} x {\rm sin} y\right]\nonumber\\
h_{32}^{AB}&=&h_{23}^{AB,*}, h_{23}^{BA}=h_{23}^{AB}, h_{32}^{BA}=h_{23}^{AB,*}\nonumber\\
h_{33}^{AB}&=&2{\rm cos} z\left[2s^{22}_{22yz}{\rm cos} 2y+(3s^{22}_{11yz}+s^{22}_{22yz}){\rm cos} x {\rm cos} y\right] +2\sum_{i=1}^{3}s^{22}_{z,2i-1}{\rm cos}(2i-1)z\nonumber\\
h_{33}^{BA}&=&h_{33}^{AB,*}=h_{33}^{AB}
\end{eqnarray}
The hopping parameters are obtained by least-square-root fitting of the above tight binding model to the DFT band structure. These parameters are given in Table \ref{hopping1} and \ref{hopping2} and the band structures of tight binding model and DFT are shown in Fig.\ref{bandTB}. Near the Fermi level, the tight binding band fits well with that of DFT.

\begin{table}[bt]%The best place to locate the table environment is directly after its first reference in text
\caption{\label{hopping1} The hopping parameters (in unit of eV) along $c$ axis to fit  the DFT results in  the six-orbital model. The onsite energies are $\epsilon_1=1.9080$eV, $\epsilon_2=2.0407$eV, $\epsilon_3=1.9722$eV and $\epsilon_4=1.9412$eV.}
\begin{ruledtabular}
\begin{tabular}{cccccccc}
%\textrm{Left\footnote{Note a.}}& \textrm{Centered\footnote{Note
%b.}}& \multicolumn{1}{c}{\textrm{Decimal}}&
%\textrm{Right}\\
$s^{\alpha\beta}_i$ & $i=z,1$ & $i=z,2$ & $i=z,3$ & $i=z,4$ & $i=z,5$ & $i=z,6$ & $i=z,7$ \\
 \colrule
$\alpha\beta$=11  & 0.1749 & 0.1568 & 0.0301 & -0.0532 & -0.0097 & -0.0189 & 0  \\
$\alpha\beta$=22  & -0.0041 & 0.1734 & 0.0019 & -0.0452 & 0 &    0    &  -0.004        \\
\end{tabular}
\end{ruledtabular}
\end{table}

\begin{table}[bt]%The best place to locate the table environment is directly after its first reference in text
\caption{\label{hopping2} The inplane hopping parameters (in unit of eV) to fit  the DFT results in  the six-orbital model. }
\begin{ruledtabular}
\begin{tabular}{cccccccc}
%\textrm{Left\footnote{Note a.}}& \textrm{Centered\footnote{Note
%b.}}& \multicolumn{1}{c}{\textrm{Decimal}}&
%\textrm{Right}\\
$s^{\alpha\beta}_i$ &  $i=1y$ & $i=2y$ & $i=11y$ & $i=22y$ & $i=12y$ & $i=y$ & $i=yz$ \\
 \colrule
$\alpha\beta$=12,$s^{12}_{i}$ & 0.0206  & 0 &  &   &   &  &        \\
$\alpha\beta$=12,$s^{12}_{iz}$ & 0.0066  & 0.0133 &   &   &  &   &          \\
$\alpha\beta$=22,$s^{22}_{i}$ &          &     & 0.0460  & -0.0191  & 0.0062 & &           \\
$\alpha\beta$=22,$s^{22}_{iz}$ &    &   & 0.0061  & -0.0109  & 0.0159 &   &        \\
$\alpha\beta$=11,$s^{11}_{i}$ &    &   &    &    &   & -0.0218 &  -0.0031  \\
\end{tabular}
\end{ruledtabular}
\end{table}
\begin{figure}[tb]
\centerline{\includegraphics[height=10 cm]{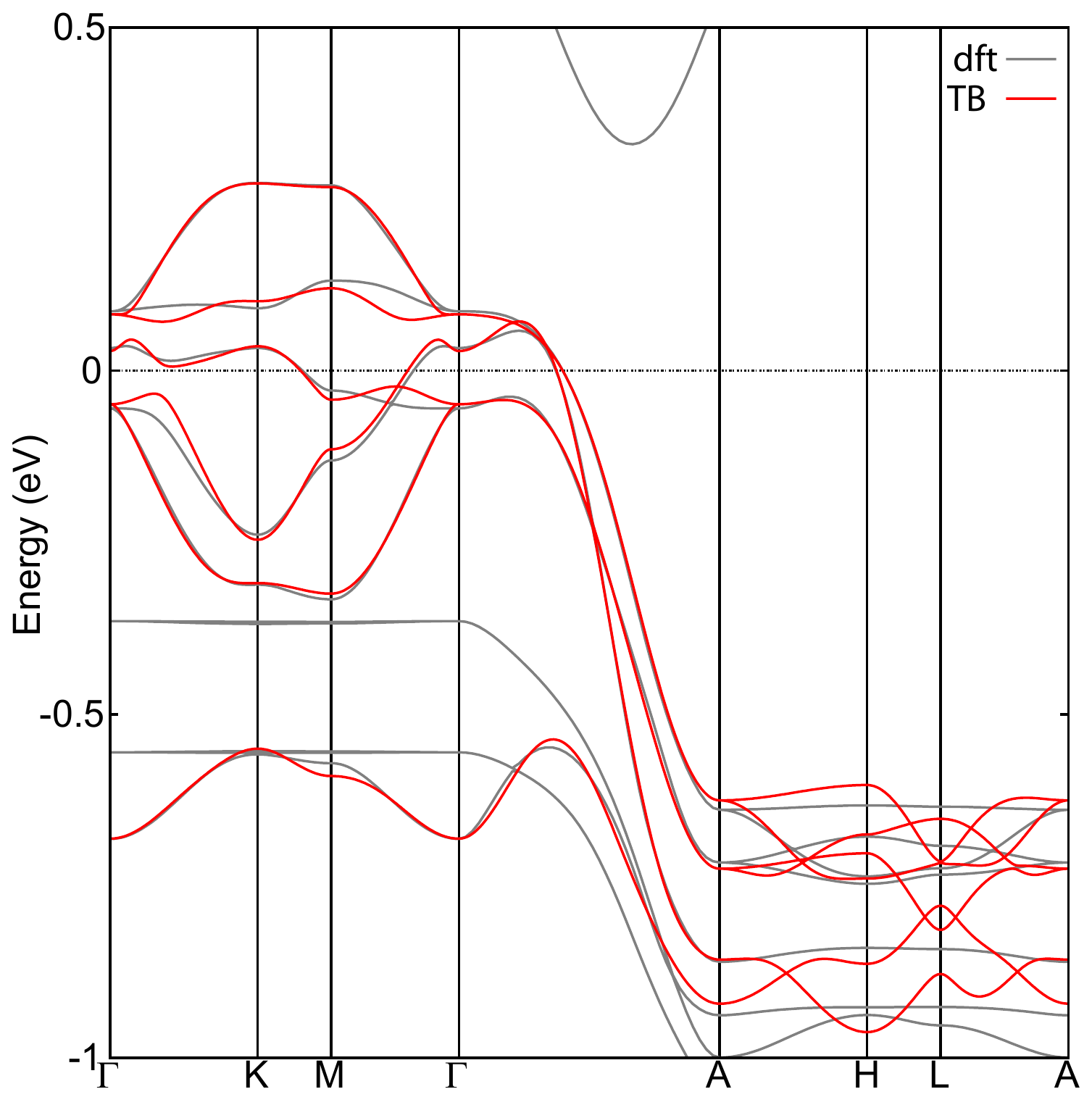}}
\caption{(color online). The band structures of DFT (gray lines) and tight binding(red lines).  \label{bandTB} }
\end{figure}

\begin{figure}[tb]
\centerline{\includegraphics[height=7 cm]{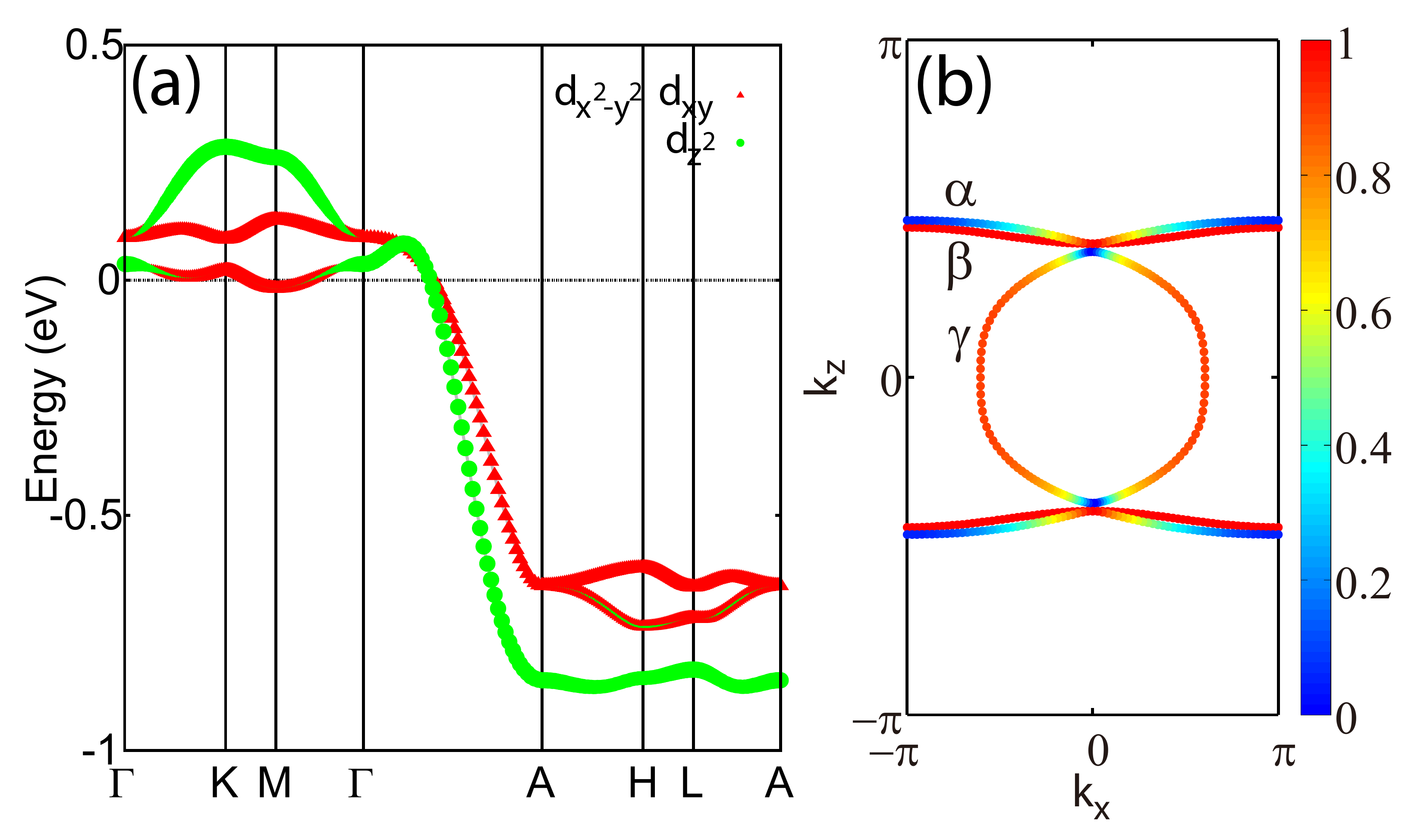}}
\caption{(color online). Orbital characters. (a) The orbital characters of the band structure of three-band tight binding model. (b) The weight distribution of $E'(d_{xy},d_{x^2-y^2})$ orbitals on Fermi surfaces in $k_y=0$ plane.   \label{character} }
\end{figure}

As the states near the Fermi level are dominantly contributed by Cr2, we can drop Cr1 and further simplify our model to a three-band one. The Hamiltonian of the resulting tight-binding model is ,
\begin{eqnarray}
h(\mathbf{k})=\left(\begin{array}{ccc}
h_{11}& h_{12} & h_{13} \\
 & h_{22} & h_{23} \\
 &               & h_{33}\\
 \end{array}\right).\label{hk}
\end{eqnarray}
Again, the unshown matrix elements are obtained by Hermicity of the $h(\mathbf{k})$ matrix. The formula of the matrix elements of $h(\mathbf{k})$ are given in Eq.(\ref{matrix}), with the hopping parameters in this model given in Table \ref{hopping3band} in the main text. The band structure of this model is shown in Fig.\ref{bandTB3} in the main text, in comparison with that of DFT. This tight binding band also fits well with that of DFT near the Fermi level. Furthermore, the orbital characters, shown in Fig.\ref{character}(a), are consistent with those of DFT\cite{Wu1}. Fig.\ref{character} shows the weight distribution of $E'$ orbitals on Fermi surfaces in $k_y=0$ plane.

\section{The multi-orbital RPA approach}

The Hamitonian adopted in our calculations is
\begin{eqnarray}
H&=&H_{\rm t-b}+H_{int}\nonumber\\
H_{int}&=&U\sum_{i\mu}n_{i\mu\uparrow}n_{i\mu\downarrow}+
V\sum_{i,\mu<\nu}n_{i\mu}n_{i\nu}+J_{H}\sum_{i,\mu<\nu}\Big[
\sum_{\sigma\sigma^{\prime}}c^{+}_{i\mu\sigma}c^{+}_{i\nu\sigma^{\prime}}
c_{i\mu\sigma^{\prime}}c_{i\nu\sigma}+(c^{+}_{i\mu\uparrow}c^{+}_{i\mu\downarrow}
c_{i\nu\downarrow}c_{i\nu\uparrow}+h.c.)\Big]\label{H-H-model_app}
\end{eqnarray}
Let's first study the case without SOC, then consider the inclusion of on-site SOC for K$_2$Cr$_3$As$_3$.
\subsection{Without SOC}
 Let's define the following bare susceptibility for the non-interacting case ($U=V=J_H=0$),
  \begin{equation}
 \chi^{(0)l_{1},l_{2}}_{l_{3},l_{4}}\left(\mathbf{q},\tau\right)\equiv
 \frac{1}{N}\sum_{\mathbf{k_{1},k_{2}}}\left<T_{\tau}c^{\dagger}_{l_{1}}(\mathbf{k_{1}},\tau)
 c_{l_{2}}(\mathbf{k_{1}+q},\tau)c^{+}_{l_{3}}(\mathbf{k_{2}+q},0)c_{l_{4}}(\mathbf{k_{2}},0)\right>_0,\label{bare}
 \end{equation}
 where $l_{i}$ $(i=1,\cdots,4)$ denote orbital indices. The explicit formalism of $\chi^{(0)}$ in the momentum-frequency space is,
\begin{equation}
 \chi^{(0)l_{1},l_{2}}_{l_{3},l_{4}}\left(\mathbf{q},i\omega_n\right)=\frac{1}{N}\sum_{\mathbf{k},\alpha,\beta}
 \xi^{\alpha}_{l_{4}}(\mathbf{k})\xi_{l_{1}}^{\alpha,*}(\mathbf{k})\xi^{\beta}_{l_{2}}(\mathbf{k+q})\xi^{\beta,*}_{l_{3}}(\mathbf{k+q})
 \frac{n_{F}(\varepsilon^{\beta}_{\mathbf{k+q}})-n_{F}(\varepsilon^{\alpha}_{\mathbf{k}})}{i\omega_n+\varepsilon^{\alpha}_{\mathbf{k}}-\varepsilon^{\beta}_{\mathbf{k+q}}},\label{explicit_free}
 \end{equation}
 where $\alpha/\beta=1,...,3$ are band indices, $\varepsilon^{\alpha}_{\mathbf{k}}$ and $\xi^{\alpha}_{l}\left(\mathbf{k}\right)$ are the $\alpha-$th eigenvalue and eigenvector
 of the $h(\mathbf{k})$ matrix respectively and $n_F$ is the Fermi-Dirac distribution
function.
\begin{figure}[tb]
\centerline{\includegraphics[height=10 cm]{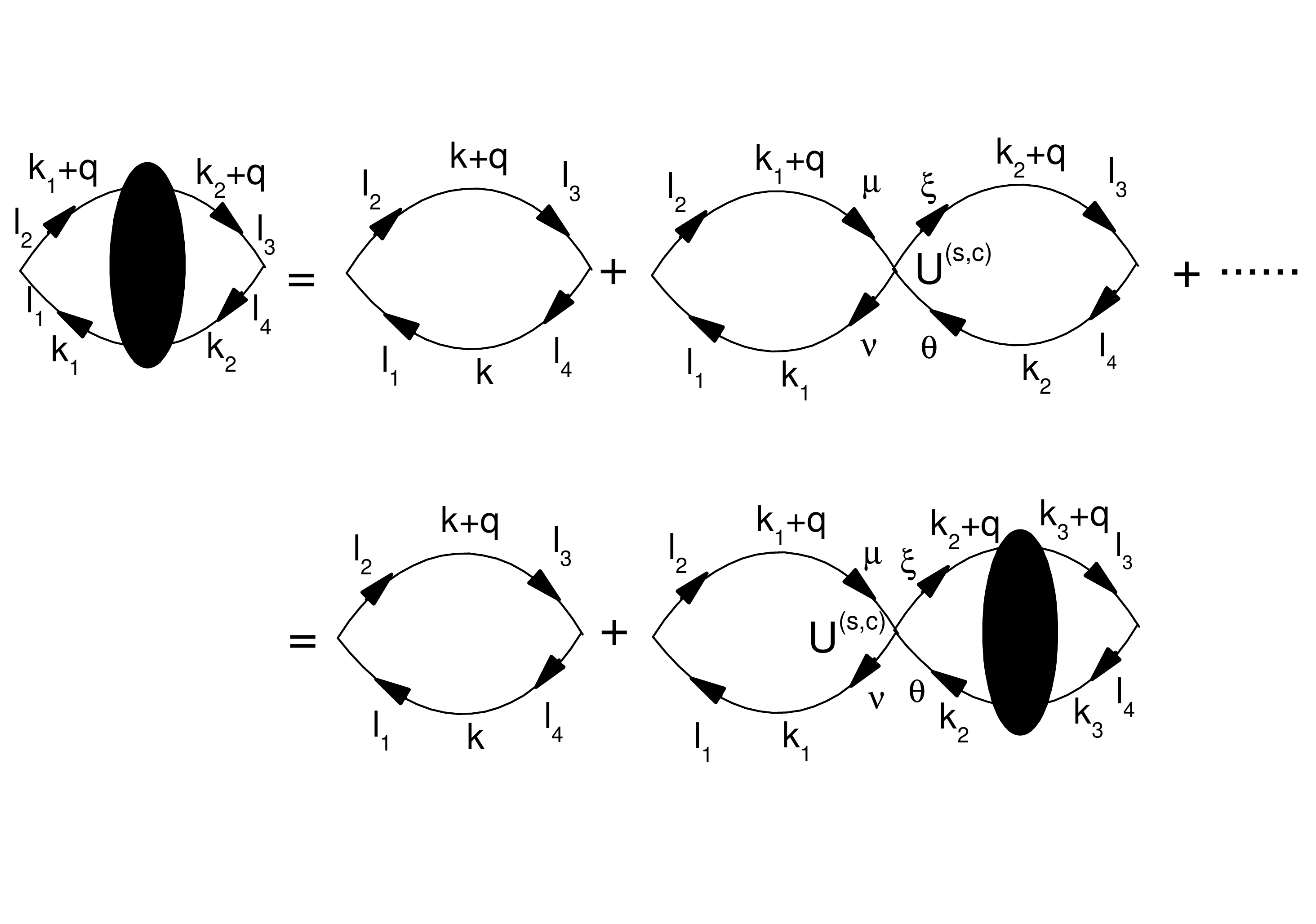}}
\caption{(color online). Feyman's diagram for the renormalized susceptibilities in the RPA level.  \label{RPA_diagram} }
\end{figure}

 When interactions turn on, we define the spin ($\chi^{(s)}$) and charge ($\chi^{(c)}$) susceptibility as follow,
\begin{eqnarray}
 \chi^{(c)l_{1},l_{2}}_{l_{3},l_{4}}\left(\mathbf{q},\tau\right)&\equiv&
 \frac{1}{2N}\sum_{\mathbf{k_{1},k_{2}},\sigma_{1},\sigma_{2}}\left<T_{\tau}C^{\dagger}_{l_{1},\sigma_{1}}(\mathbf{k_{1}},\tau)
 C_{l_{2},\sigma_{1}}(\mathbf{k_{1}+q},\tau)C^{+}_{l_{3},\sigma_{2}}(\mathbf{k_{2}+q},0)C_{l_{4},\sigma_{2}}(\mathbf{k_{2}},0)\right>,
 \nonumber\\
 \chi^{\left(s^{z}\right)l_{1},l_{2}}_{l_{3},l_{4}}\left(\mathbf{q},\tau\right)&\equiv&
 \frac{1}{2N}\sum_{\mathbf{k_{1},k_{2}},\sigma_{1},\sigma_{2}}\sigma_{1}\sigma_{2}\left<T_{\tau}C^{\dagger}_{l_{1},\sigma_{1}}(\mathbf{k_{1}},\tau)
 C_{l_{2},\sigma_{1}}(\mathbf{k_{1}+q},\tau)C^{+}_{l_{3},\sigma_{2}}(\mathbf{k_{2}+q},0)C_{l_{4},\sigma_{2}}(\mathbf{k_{2}},0)\right>,\nonumber\\
 \chi^{\left(s^{+-}\right)l_{1},l_{2}}_{l_{3},l_{4}}\left(\mathbf{q},\tau\right)&\equiv&
 \frac{1}{N}\sum_{\mathbf{k_{1},k_{2}}}\left<T_{\tau}C^{\dagger}_{l_{1}\uparrow}(\mathbf{k_{1}},\tau)
 C_{l_{2}\downarrow}(\mathbf{k_{1}+q},\tau)C^{+}_{l_{3}\downarrow}(\mathbf{k_{2}+q},0)C_{l_{4}\uparrow}(\mathbf{k_{2}},0)\right>,\nonumber\\
  \chi^{\left(s^{-+}\right)l_{1},l_{2}}_{l_{3},l_{4}}\left(\mathbf{q},\tau\right)&\equiv&
 \frac{1}{N}\sum_{\mathbf{k_{1},k_{2}}}\left<T_{\tau}C^{\dagger}_{l_{1}\downarrow}(\mathbf{k_{1}},\tau)
 C_{l_{2}\uparrow}(\mathbf{k_{1}+q},\tau)C^{+}_{l_{3}\uparrow}(\mathbf{k_{2}+q},0)C_{l_{4}\downarrow}(\mathbf{k_{2}},0)\right>.
 \end{eqnarray}
 Note that in non-magnetic state we have
 $\chi^{\left(s^{z}\right)}=\chi^{\left(s^{+-}\right)}=\chi^{\left(s^{-+}\right)}\equiv\chi^{\left(s\right)}$, and when $U=V=J_H=0$ we have $\chi^{(c)}=\chi^{(s)}=\chi^{(0)}$.
 In the RPA level, the renormalized spin/charge susceptibilities for the system are,
 \begin{eqnarray}
\chi^{(s)}\left(\mathbf{q},i\nu\right)&=&\left[I-\chi^{(0)}\left(\mathbf{q},i\nu\right)U^{(s)}\right]^{-1}\chi^{(0)}\left(\mathbf{q},i\nu\right),\nonumber\\
\chi^{(c)}\left(\mathbf{q},i\nu\right)&=&\left[I+\chi^{(0)}\left(\mathbf{q},i\nu\right)U^{(c)}\right]^{-1}\chi^{(0)}\left(\mathbf{q},i\nu\right),\label{RPA_SUS}
\end{eqnarray}
where $\chi^{(s,c)}\left(\mathbf{q},i\nu_{n}\right)$, $\chi^{(0)}\left(\mathbf{q},i\nu_{n}\right)$ and $U^{(s,c)}$ are operated as
$9\times 9$ matrices (the upper or lower two indices are viewed as one number) with elements of the matrix $U^{(s,c)}$ to be
\begin{eqnarray}
U^{(s) l_{1}l_{2}}_{l_{3}l_{4}}&=&\left\{\begin{array}{cc}{U,l_{1}=l_{2}=l_{3}=l_{4}}\\{J_H,l_{1}=l_{2}\ne
l_{3}=l_{4}}\\{J_{H},l_{1}=l_{3}\ne
l_{2}=l_{4}}\\{V,l_{1}=l_{4}\ne
l_{3}=l_{2}}\end{array}\right.
\end{eqnarray}

\begin{eqnarray}
U^{(c) l_{1}l_{2}}_{l_{3}l_{4}}&=&\left\{\begin{array}{cc}{U,l_{1}=l_{2}=l_{3}=l_{4}}\\{2V-J_H,l_{1}=l_{2}\ne
l_{3}=l_{4}}\\{J_{H},l_{1}=l_{3}\ne
l_{2}=l_{4}}\\{2J_{H}-V,l_{1}=l_{4}\ne
l_{3}=l_{2}}\end{array}\right.
\end{eqnarray}
The Feyman's diagram of RPA is shown in Fig.\ref{RPA_diagram} For repulsive Hubbard-interactions, the spin susceptibility is enhanced and the charge susceptibility is suppressed. Note that there is a critical interaction strength $U_c$ which depends on the ratio $J_H/U$. When $U>U_c$, the denominator
matrix $I-\chi^{(0)}\left(\mathbf{q},i\nu\right)U^{(s)}$ in Eq.(\ref{RPA_SUS}) will have zero eigenvalues for some $\mathbf{q}$ and the renormalized spin susceptibility diverges there, which invalidates the RPA treatment. This divergence of spin susceptibility for $U>U_c$ implies magnetic order. When $U<U_c$, the short-ranged spin or charge fluctuations would mediate Cooper pairing in the system.

\begin{figure}
\scalebox{0.4}{{\includegraphics[scale=0.55]{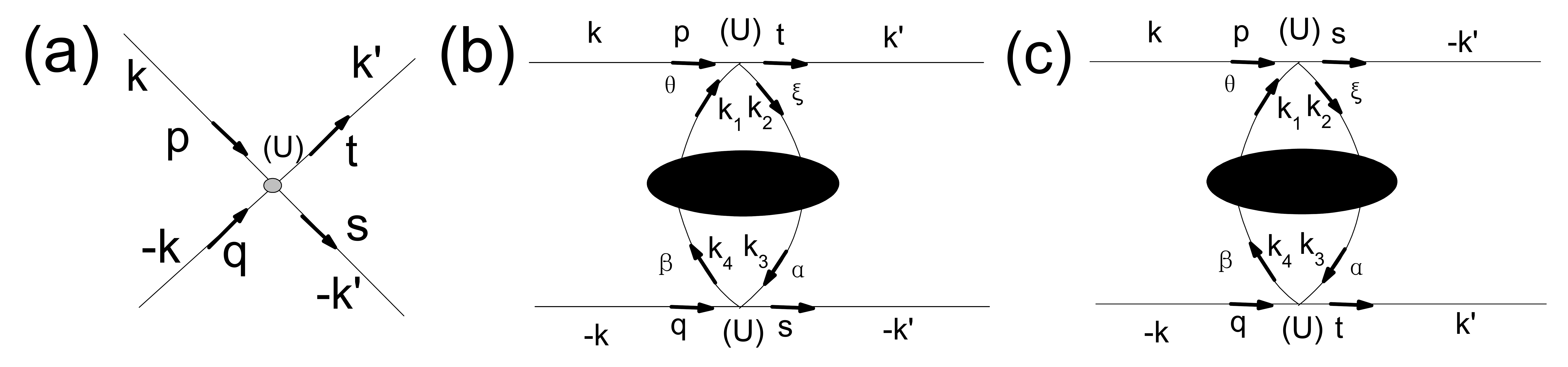}}
}\caption{The three processes which contribute the renormalized effective vertex considered in the RPA, with (a) the bare interation vertex and (b) (c) the two second order perturbation processes during which spin or charge fluctuations are exchanged between a cooper pair.} \label{RPA_effective}
\end{figure}

 Let's consider a Cooper pair with momentum/orbital $(\mathbf{k'}t,-\mathbf{k'}s)$, which could be scattered to
 $(\mathbf{k}p,-\mathbf{k}q)$ by exchanging charge or spin fluctuations. In the RPA level, The effective interaction induced by this process is as follow,
 \begin{equation}
 V^{RPA}_{eff}=\frac{1}{N}\sum_{pqst,kk'}\Gamma^{pq}_{st}(k,k')c_{p}^{\dagger}
 (\mathbf{k})c_{q}^{\dagger}(-\mathbf{k})c_{s}(-\mathbf{k}')c_{t}(\mathbf{k}'),\label{effective_vertex}
 \end{equation}
We consider the three processes in Fig.\ref{RPA_effective} which contribute to the effective vertex $\Gamma^{pq}_{st}(k,k')$, where (a) represents the bare interaction vertex and (b) (c) represent the two second order perturbation processes during which spin or charge fluctuations are exchanged between a cooper pair. In the singlet channel, the effective vertex $\Gamma^{pq}_{st}(k,k')$ is given as follow,
\begin{eqnarray}
\Gamma^{pq(s)}_{st}(k,k')=\left(\frac{U^{(c)}+3U^{(s)}}{4}\right)^{pt}_{qs}+
\frac{1}{4}\left[3U^{(s)}\chi^{(s)}\left(k-k'\right)U^{(s)}-U^{(c)}\chi^{(c)}\left(k-k'\right)U^{(c)}\right]^{pt}_{qs}+\nonumber\\
\frac{1}{4}\left[3U^{(s)}\chi^{(s)}\left(k+k'\right)U^{(s)}-U^{(c)}\chi^{(c)}\left(k+k'\right)U^{(c)}\right]^{ps}_{qt},
 \end{eqnarray}
 while in the triplet channel, it is
\begin{eqnarray}
\Gamma^{pq(t)}_{st}(k,k')=\left(\frac{U^{(c)}-U^{(s)}}{4}\right)^{pt}_{qs}-
\frac{1}{4}\left[U^{(s)}\chi^{(s)}\left(k-k'\right)U^{(s)}+U^{(c)}\chi^{(c)}\left(k-k'\right)U^{(c)}\right]^{pt}_{qs}+\nonumber\\
\frac{1}{4}\left[U^{(s)}\chi^{(s)}\left(k+k'\right)U^{(s)}+U^{(c)}\chi^{(c)}\left(k+k'\right)U^{(c)}\right]^{ps}_{qt},
 \end{eqnarray}
   Notice that the vertex $\Gamma^{pq}_{st}(k,k')$ has been symmetrized for the singlet case and anti-symmetrized for the
 triplet case. Generally we neglect the frequency-dependence of $\Gamma$ and replace it by $\Gamma^{pq}_{st}(k,k')\approx\Gamma^{pq}_{st}(\mathbf{k,k'},0)$. Usually, only the real part of $\Gamma$ is adopted\cite{Scalapino1,Scalapino2}

 Considering only intra-band pairings, we obtain the following effective pairing interaction on the FSs,
 \begin{eqnarray}
 V_{eff}=
 \frac{1}{N}\sum_{\alpha\beta,\mathbf{k}\mathbf{k'}}V^{\alpha\beta}(\mathbf{k,k'})c_{\alpha}^{\dagger}(\mathbf{k})
 c_{\alpha}^{\dagger}(-\mathbf{k})c_{\beta}(-\mathbf{k}')c_{\beta}(\mathbf{k}'),\label{pairing_interaction}
 \end{eqnarray}
where $\alpha/\beta=1,\cdots,3$ are band indices and $V^{\alpha\beta}(\mathbf{k,k'})$ is
 \begin{eqnarray}
 V^{\alpha\beta}(\mathbf{k,k'})=\sum_{pqst,\mathbf{k}\mathbf{k'}}\Gamma^{pq}_{st}(\mathbf{k,k'},0)\xi_{p}^{\alpha,*}(\mathbf{k})
 \xi_{q}^{\alpha,*}(-\mathbf{k})\xi_{s}^{\beta}(-\mathbf{k'})\xi_{t}^{\beta}(\mathbf{k'}).\label{effective_potential}
 \end{eqnarray}
 From the effective pairing interaction (\ref{pairing_interaction}), one can obtain the following linearized gap
 equation\cite{Scalapino1,Scalapino2} to determine the $T_c$ and the leading pairing symmetry of the system,
  \begin{equation}
 -\frac{1}{(2\pi)^3}\sum_{\beta}\iint_{FS}
d^{2}\mathbf{k'}_{\Vert}\frac{V^{\alpha\beta}(\mathbf{k,k'})}{v^{\beta}_{F}(\mathbf{k'})}\Delta_{\beta}(\mathbf{k'})=\lambda
 \Delta_{\alpha}(\mathbf{k}).\label{eigenvalue_Tc}
\end{equation}
 This equation can be looked upon as an eigenvalue problem, where the normalized eigenvector $\Delta_{\alpha}(\mathbf{k})$ represents the relative gap function on the $\alpha-$th FS patches near $T_c$, and the eigenvalue $\lambda$ determines $T_c$ via $T_{c}$=cut off energy $e^{-1/\lambda}$, where the cut off energy is of order band-width. The leading pairing symmetry is determined by the largest eigenvalue $\lambda$ of Eq.(\ref{eigenvalue_Tc}).

 \subsection {With on-site SOC}
 When the on-site SOC described by Eq.(\ref{SOC}) turns on, the free electron part of our model becomes,
 \begin{equation}
H_{\rm band}=H_{\rm TB}+H_{\rm SOC}.\label{band_with_soc}
\end{equation}
Let's rewrite the interaction part in the following formalism,
\begin{equation}
H_{\rm int}=\sum_{i\mu\nu\theta\xi\sigma_{1}\sigma_{2}}U
^{\mu\nu\sigma_{1}}_{\theta\xi\sigma_{2}}c^{\dagger}_{i\mu\sigma_{1}}
c_{i\nu\sigma_{1}}c^{\dagger}_{i\theta\sigma_{2}}c_{i\xi\sigma_{2}}.\label{h_int}
\end{equation}
Here $U ^{\mu\nu\sigma_{1}}_{\theta\xi\sigma_{2}}$ can be defined to satisfy the following symmetry relation,
\begin{eqnarray}
U^{\mu\nu\uparrow}_{\theta\xi\uparrow}=U^{\mu\nu\downarrow}_{\theta\xi\downarrow}\equiv
\bar{U}^{\mu\nu}_{\theta\xi}\nonumber\\
U^{\mu\nu\uparrow}_{\theta\xi\downarrow}=U^{\mu\nu\downarrow}_{\theta\xi\uparrow}\equiv
\bar{V} ^{\mu\nu}_{\theta\xi},\label{symmetry_1}
\end{eqnarray}
where the $\bar{U},\bar{V}$ tensors are obtained as
\begin{eqnarray}
\bar{U}^{l_{1}l_{2}}_{l_{3}l_{4}}&=&\left\{\begin{array}{cc}{\frac{U_{2}-J_{H}}{4},l_{1}=l_{2}\ne
l_{3}=l_{4}}\\{\frac{J_{H}-U_{2}}{4},l_{1}=l_{4}\ne
l_{3}=l_{2}}\end{array}\right.\nonumber\\\bar{V}^{l_{1}l_{2}}_{l_{3}l_{4}}&=&\left\{\begin{array}{cc}{\frac{U_{1}}{2},l_{1}=l_{2}=l_{3}=l_{4}}\\{\frac{U_{2}}{2},l_{1}=l_{2}\ne
l_{3}=l_{4}}\\{\frac{J_{H}}{2},l_{1}=l_{3}\ne
l_{2}=l_{4}}\\{\frac{J_{H}}{2},l_{1}=l_{4}\ne
l_{3}=l_{2}}\end{array}\right.
\end{eqnarray}

Due to the $S_z$-conservation of the on-site SOC, let's define the following six susceptibilities,
 \begin{eqnarray}
 \chi^{1}_{l_{1},l_{2},l_{3},l_{4}}\left(\mathbf{q},\tau\right)&\equiv&
 \frac{1}{N}\sum_{\mathbf{k_{1},k_{2}}}\left<T_{\tau}c^{\dagger}_{l_{1}\uparrow}(\mathbf{k_{1}},\tau)
 c_{l_{2}\uparrow}(\mathbf{k_{1}+q},\tau)c^{+}_{l_{3}\uparrow}(\mathbf{k_{2}+q},0)c_{l_{4}\uparrow}(\mathbf{k_{2}},0)\right>\nonumber\\
\chi^{2}_{l_{1},l_{2},l_{3},l_{4}}\left(\mathbf{q},\tau\right)&\equiv&
 \frac{1}{N}\sum_{\mathbf{k_{1},k_{2}}}\left<T_{\tau}c^{\dagger}_{l_{1}\uparrow}(\mathbf{k_{1}},\tau)
 c_{l_{2}\uparrow}(\mathbf{k_{1}+q},\tau)c^{+}_{l_{3}\downarrow}(\mathbf{k_{2}+q},0)c_{l_{4}\downarrow}(\mathbf{k_{2}},0)\right>\nonumber\\
 \chi^{3}_{l_{1},l_{2},l_{3},l_{4}}\left(\mathbf{q},\tau\right)&\equiv&
 \frac{1}{N}\sum_{\mathbf{k_{1},k_{2}}}\left<T_{\tau}c^{\dagger}_{l_{1}\downarrow}(\mathbf{k_{1}},\tau)
 c_{l_{2}\downarrow}(\mathbf{k_{1}+q},\tau)c^{+}_{l_{3}\uparrow}(\mathbf{k_{2}+q},0)c_{l_{4}\uparrow}(\mathbf{k_{2}},0)\right>\nonumber\\
 \chi^{4}_{l_{1},l_{2},l_{3},l_{4}}\left(\mathbf{q},\tau\right)&\equiv&
 \frac{1}{N}\sum_{\mathbf{k_{1},k_{2}}}\left<T_{\tau}c^{\dagger}_{l_{1}\downarrow}(\mathbf{k_{1}},\tau)
 c_{l_{2}\downarrow}(\mathbf{k_{1}+q},\tau)c^{+}_{l_{3}\downarrow}(\mathbf{k_{2}+q},0)c_{l_{4}\downarrow}(\mathbf{k_{2}},0)\right>\nonumber\\
 \chi^{5}_{l_{1},l_{2},l_{3},l_{4}}\left(\mathbf{q},\tau\right)&\equiv&
 \frac{1}{N}\sum_{\mathbf{k_{1},k_{2}}}\left<T_{\tau}c^{\dagger}_{l_{1}\uparrow}(\mathbf{k_{1}},\tau)
 c_{l_{2}\downarrow}(\mathbf{k_{1}+q},\tau)c^{+}_{l_{3}\downarrow}(\mathbf{k_{2}+q},0)c_{l_{4}\uparrow}(\mathbf{k_{2}},0)\right>\nonumber\\
 \chi^{6}_{l_{1},l_{2},l_{3},l_{4}}\left(\mathbf{q},\tau\right)&\equiv&
 \frac{1}{N}\sum_{\mathbf{k_{1},k_{2}}}\left<T_{\tau}c^{\dagger}_{l_{1}\downarrow}(\mathbf{k_{1}},\tau)
 c_{l_{2}\uparrow}(\mathbf{k_{1}+q},\tau)c^{+}_{l_{3}\uparrow}(\mathbf{k_{2}+q},0)c_{l_{4}\downarrow}(\mathbf{k_{2}},0)\right>
 \end{eqnarray}
 For $(U,V,J_{H})=(0,0,0)$, we find the following formula of the bare susceptibilities,
 \begin{eqnarray}
 \chi^{1(0)}_{l_{1},l_{2},l_{3},l_{4}}\left(\mathbf{q},i\omega_{n}\right)&=&
 \frac{1}{N}\sum_{\mathbf{k},i\omega_m}G^{\uparrow(0)}_{l_{4},l_{1}}\left(\mathbf{k},i\omega_{m}\right)G^{\uparrow(0)}_{l_{2},l_{3}}\left(\mathbf{k+q},i\omega_{m}+i\omega_{n}\right)
 \nonumber\\&=&\frac{1}{N}\sum_{\mathbf{k},\alpha,\beta}
 \xi^{\alpha}_{l_{4}\uparrow}(\mathbf{k})\xi_{l_{1}\uparrow}^{\alpha,*}(\mathbf{k})\xi^{\beta}_{l_{2}\uparrow}(\mathbf{k+q})\xi^{\beta,*}_{l_{3}\uparrow}(\mathbf{k+q})
 \frac{n_{F}(\varepsilon^{\beta}_{\mathbf{k+q}})-n_{F}(\varepsilon^{\alpha}_{\mathbf{k}})}
 {i\omega_n+\varepsilon^{\alpha}_{\mathbf{k}}-\varepsilon^{\beta}_{\mathbf{k+q}}}\nonumber\\
  \chi^{2(0)}_{l_{1},l_{2},l_{3},l_{4}}\left(\mathbf{q},i\omega_{n}\right)&=&\chi^{3(0)}_{l_{1},l_{2},l_{3},l_{4}}=0\nonumber\\
  \chi^{4(0)}_{l_{1},l_{2},l_{3},l_{4}}(\mathbf{q},i\omega_{n})&=&
 \chi^{1(0),*}_{l_{1},l_{2},l_{3},l_{4}}(\mathbf{q},-i\omega_{n})\nonumber\\
  \chi^{6(0)}_{l_{1},l_{2},l_{3},l_{4}}\left(\mathbf{q},i\omega_{n}\right)&=&
 \frac{1}{N}\sum_{\mathbf{k},i\omega_m}G^{\downarrow(0)}_{l_{4},l_{1}}\left(\mathbf{k},i\omega_{m}\right)G^{\uparrow(0)}_{l_{2},l_{3}}\left(\mathbf{k+q},i\omega_{m}+i\omega_{n}\right)
 \nonumber\\&=&\frac{1}{N}\sum_{\mathbf{k},\alpha,\beta}
 \xi^{\alpha,*}_{l_{4}\downarrow}(\mathbf{k})\xi_{l_{1}\downarrow}^{\alpha}(\mathbf{k})\xi^{\beta}_{l_{2}\uparrow}(\mathbf{k+q})\xi^{\beta,*}_{l_{3}\uparrow}(\mathbf{k+q})
 \frac{n_{F}(\varepsilon^{\beta}_{\mathbf{k+q}})-n_{F}(\varepsilon^{\alpha}_{\mathbf{k}})}
 {i\omega_n+\varepsilon^{\alpha}_{\mathbf{k}}-\varepsilon^{\beta}_{\mathbf{k+q}}}\nonumber\\
 \chi^{5(0)}_{l_{1},l_{2},l_{3},l_{4}}(\mathbf{q},i\omega_{n})&=&
\chi^{6(0),*}_{l_{1},l_{2},l_{3},l_{4}}(\mathbf{q},-i\omega_{n})
\end{eqnarray}

 When interactions turn on, we obtain the following Dyson equations for these susceptibilities in the RPA level, represented by the Feyman's diagram shown in Fig.{\ref{RPA_diagram}}
\begin{eqnarray}
 \chi^{1}&=&\chi^{1(0)}-\chi^{1(0)}\left(4\bar{U}\chi^{1}+2\bar{V}\chi^{3}\right)\nonumber\\
  \chi^{2}&=&-\chi^{1(0)}\left(4\bar{U}\chi^{2}+2\bar{V}\chi^{4}\right)\nonumber\\
 \chi^{3}&=&-\chi^{4(0)}\left(4\bar{U}\chi^{3}+2\bar{V}\chi^{1}\right)\nonumber\\
\chi^{4}&=&\chi^{4(0)}-\chi^{4(0)}\left(4\bar{U}\chi^{4}+2\bar{V}\chi^{2}\right)\nonumber\\
\chi^{5}&=&\chi^{5(0)}-\chi^{5(0)}\left(4\bar{U}-2\bar{V}\right)\chi^{5}\nonumber\\
 \chi^{6}&=&\chi^{6(0)}-\chi^{6(0)}\left(4\bar{U}-2\bar{V}\right)\chi^{6}.\label{dyson}
 \end{eqnarray}
Here $\bar{U}$ and $\bar{V}$ are operated as matrices with the upper or lower two tensor indices taken as one number. These equations are easily solved.

The effective interaction vertex $\Gamma^{pq\s_1}_{st\s_2}(\mathbf{k,k'})$ is obtained from the summation of the three processes listed in Fig.\ref{RPA_effective},
\begin{equation}
\Gamma^{pq\s_1}_{st\s_2}(\mathbf{k,k'})=\Gamma^{(a)pq\s_1}_{st\s_2}(\mathbf{k,k'})+\Gamma^{(b)pq\s_1}_{st\s_2}(\mathbf{k,k'})+\Gamma^{(c)pq\s_1}_{st\s_2}(\mathbf{k,k'}).\label{sum}
\end{equation}
Firstly, the contribution from (a) is,
\begin{equation}
\Gamma^{(a)pq\sigma_{1}}_{st\sigma_{2}}(\mathbf{k,k'})=(U)^{pt\sigma_{1}}_{qs\sigma_{2}}.\label{bare}
 \end{equation}
Then the contribution from (b) and (c) for $\uparrow\uparrow$ and $\uparrow\downarrow$ is,
\begin{eqnarray}
 \Gamma^{(b)pq\uparrow}_{st\downarrow}(\mathbf{k,k'})&=&-\left[16\bar{U}\chi^{2}\left(\mathbf{k}-\mathbf{k'}\right)\bar{U}+4\bar{V}\chi^{3}\left(\mathbf{k}-\mathbf{k'}\right)\bar{V}
 +8\bar{U}\chi^{1}\left(\mathbf{k}-\mathbf{k'}\right)\bar{V}+8\bar{V}\chi^{4}\left(\mathbf{k}-\mathbf{k'}\right)\bar{U}\right]^{pt}_{qs}\nonumber\\
 \Gamma^{(b)pq\uparrow}_{st\uparrow}(\mathbf{k,k'})&=&-\left[16\bar{U}\chi^{1}\left(\mathbf{k}-\mathbf{k'}\right)\bar{U}+8\bar{U}\chi^{2}\left(\mathbf{k}-\mathbf{k'}\right)\bar{V}
 +8\bar{V}\chi^{3}\left(\mathbf{k}-\mathbf{k'}\right)\bar{U}+4\bar{V}\chi^{4}\left(\mathbf{k}-\mathbf{k'}\right)\bar{V}\right]^{pt}_{qs}\nonumber\\
\Gamma^{(c)pq\uparrow}_{st\downarrow}(\mathbf{k,k'})&=&4\left[\left(\bar{V}-2\bar{U}\right)\chi^{6}\left(\mathbf{k}+\mathbf{k'}\right)\left(\bar{V}-2\bar{U}\right)\right]^{ps}_{qt}.\nonumber\\
\Gamma^{(c)pq\uparrow}_{st\uparrow}(\mathbf{k,k'})&=&\left[16\bar{U}\chi^{1}\left(\mathbf{k}+\mathbf{k'}\right)\bar{U}+8\bar{U}\chi^{2}\left(\mathbf{k}+\mathbf{k'}\right)\bar{V}
 +8\bar{V}\chi^{3}\left(\mathbf{k}+\mathbf{k'}\right)\bar{U}+4\bar{V}\chi^{4}\left(\mathbf{k}+\mathbf{k'}\right)\bar{V}\right]^{ps}_{qt}.
 \end{eqnarray}
The $\Gamma^{pq\sigma_{1}}_{st\sigma_{2}}(\mathbf{k,k'})$ for other spin configurations can be obtained by the relation
$\Gamma^{pq\sigma_{1}}_{st\sigma_{2}}(\mathbf{k,k'})=\left[\Gamma^{pq\bar{\sigma_{1}}}_{st\bar{\sigma_{2}}}(\mathbf{-k,-k'})\right]^{*}$ from time-reversal symmetry.

Finally, after the effective pairing potential  $V^{\alpha\beta}_{\s_1\s_2}(\mathbf{k,k'})$ is obtained from $\Gamma^{pq\sigma_{1}}_{st\sigma_{2}}(\mathbf{k,k'})$ via the relation (\ref{effective_potential}), we should solve the linearized gap equation (\ref{eigenvalue_Tc}) for each spin configuration $\s_1\s_2$.

\end{widetext}

\begin{references}

\bibitem{SrRuO1} Y. Maeno, H. Hashimoto, K. Yoshida, S. Nishizaki, T. Fujita, J. G. Bednorz, and F. Lichtenberg, Nature {\bf 372}, 532 (1994).

\bibitem{SrRuO2} A. P. Mackenzie and Y. Maeno, Rev. Mod. Phys. 75, 657(2003).


\bibitem{Jerome1980} D. Jerome, A. Mazaud, M. Ribault, and K. Bechgaard,
Journal De Physique Lettres {\bf 41}, L95 (1980).

\bibitem{Wilhelm2001} H. Wilhelm, D. Jaccard, R. Duprat, C. Bourbonnais, D. J��rome, J. Moser, C. Carcel and J. M. Fabre, Eur. Phys. J. B {\bf 21}, 175 (2001).



\bibitem{Armici1980} J. C. Armici, M. Decroux, O. Fischer, M. Potel, R. Chevrel, and M. Sergent,
Solid State Commun. {\bf 33}, 607 (1980).


\bibitem{Greenblatt1984} M. Greenblatt, W. H. McCarroll, R. Neifeld, M. Croft, and J. V. Waszczak,
Solid State Commun. {\bf 51}, 671 (1984).

\bibitem{Denlinger1999} J. D. Denlinger, G. H. Gweon, J. W. Allen, C. G. Olson, J. Marcus, C. Schlenker, and L. S. Hsu, Phys. Rev. Lett. {\bf 82}, 2540 (1999).
\bibitem{Xu2009} X. Xu, A. F. Bangura, J. G. Analytis, J. D. Fletcher, M. M. J. French, N. Shannon, J. He, S. Zhang, D. Mandrus, R. Jin, and N. E. Hussey, Phys. Rev. Lett. {\bf 102}, 206602 (2009).
\bibitem{Mercure2012} J. F. Mercure, A. F. Bangura, X. Xu, N. Wakeham, A. Carrington, P. Walmsley, M. Greenblatt, and N. E. Hussey, Phys. Rev. Lett. {\bf 108}, 187003 (2012).

\bibitem{KCrAs} J.-K. Bao, J.-Y. Liu, C.-W. Ma, Z.-H. Meng, Z.-T. Tang, Y.-L. Sun, H.-F. Zhai, H. Jiang, H. Bai, C.-M. Feng, Z.-A. Xu and G.-H. Cao,
                Phys. Rev. X {\bf 5}, 011013 (2015).

\bibitem{RbCrAs} Z. T. Tang, J. K. Bao, Y. Liu, Y. L. Sun, A. Ablimit, H. F. Zhai, H. Jiang, C. M. Feng, Z. A. Xu and G. H. Cao, Phys. Rev. B {\bf 91}, 020506(R) (2015).

\bibitem{CsCrAs} Z.-T. Tang, J.-K. Bao, Z. Wang, H. Bai, H. Jiang, Y. Liu, H.-F. Zhai, C.-M. Feng, Z.-A. Xu and G.-H. Cao, Science China Materials, {\bf 58}(1), 16-10 (2015)
\bibitem{Luo} W. Wu, J. Cheng, K. Matsubayashi, P. Kong, F. Lin, C. Jin, N. Wang, Y. Uwatoko and J. Luo, Nature Communications {\bf 5}, 5508 (2014).

\bibitem{Kotegawa} H. Kotegawa, S. Nakahara, H. Tou and H. Sugawara, J. Phys. Soc. Jpn. {\bf 83}, 093702 (2014).

\bibitem{Zhaojun} Y. Shen, Q. Wang, Y. Hao, B. Pan, Y. Feng, Q. Huang, L. W. Harriger, J. B. Leao, Y. Zhao, R. M. Chisnell, J. W. Lynn, H. Cao, J. Hu and J. Zhao, arXiv:1409.6615 (2014).

\bibitem{NMR_CrAs} H. Kotegawa, S. Nakahara, R. Akamatsu, H. Tou, H. Sugawara and H. Harima,  Phys. Rev. Lett. {\bf 114}, 117002 (2015).

\bibitem{NMR_KCrAs} H.- Z. Zhi, T. Imai, F. L. Ning, J.-K. Bao and G.-H. Cao,  Phys. Rev. Lett. {\bf 114}, 147004 (2015).

\bibitem{pen_depth} G. M. Pang, M. Smidman, W. B. Jiang, J. K. Bao, Z. F. Weng, Y. F. Wang, L. Jiao, J. L. Zhang, G. H. Cao and H. Q. Yuan,  Phys. Rev. B {\bf 91}, 220502(R) (2015).

\bibitem{Cao} H. Jiang, G. Cao and C. Cao,  arXiv:1412.1309.

\bibitem{Wu1} X. Wu, C. Le, J. Yuan, H. Fan and J. Hu, Chin. Phys. Lett. {\bf 32}, 057401(2015).

\bibitem{RPA1} T. Takimoto, T. Hotta, and K. Ueda, Phys. Rev. B \textbf{69}, 104504 (2004).

\bibitem{RPA2} K. Yada and H. Kontani, J. Phys. Soc. Jpn. \textbf{74}, 2161 (2005).

\bibitem{RPA3} K. Kubo, Phys. Rev. B \textbf{75}, 224509 (2007).

\bibitem{Kuroki} K. Kuroki, S. Onari, R. Arita, H. Usui, Y. Tanaka, H. Kontani and H. Aoki, Phys. Rev. Lett. \textbf{101}, 087004 (2008).

\bibitem{Scalapino1} S. Graser, T. A. Maier, P. J. Hirschfeld and D. J. Scalapino, New Journal of Physics \textbf{11}, 025016 (2009).

\bibitem{Scalapino2} T.A. Maier, S. Graser, P.J. Hirschfeld and D.J. Scalapino, Phys. Rev. B {\bf 83}, 100515(R)(2011).

\bibitem{Yang} F. Liu, C.-C. Liu, K. Wu, F. Yang and Y. Yao, Phys. Rev. Lett. \textbf{111}, 066804 (2013).
\bibitem{Wu2014}  X. Wu, J. Yuan, Y. Liang, H. Fan and J. Hu, Europhys. Lett. {\bf 108} 27006 (2014).

\bibitem{Ma2014}    T. Ma, F. Yang, H. Yao, H. Lin,   Phys. Rev. B {\bf 90}, 245114 (2014).
\bibitem{Zhang2015} L. Zhang, F. Yang, Y. Yao,   Sci. Rep. {\bf 5}, 8203 (2015).

\bibitem{BiH} Note that similar Hund's rule coupling driven on-site inter-orbital odd-parity pairing was also proposed in BiH in F. Yang, C.-C. Liu, Y.-Z. Zhang, Y. Yao, D.-H. Lee, Phys. Rev. B {\bf 91}, 134514 (2015), where similar mean-field decoupling on the Hubbard-Hund's interaction was performed.


\bibitem{Hu2012} Similar physics has been clarified in J. P. Hu and H. Ding, Sci. Rep. {\bf 2}, 381 (2012).
\bibitem{Frigeri2004} P. A. Frigeri, D. F. Agterberg and M. Sigrist, New J. Phys. {\bf 6}, 115 (2004).
\bibitem{Ishida1998} K. Ishida, H. Mukuda, Y. Kitaoka, K. Asayama, Z. Q. Mao, Y. Mori and Y. Maeno, Nature {\bf 396}, 658 (1998).
\bibitem{Zhou2015} Y. Zhou, C. Cao and F. C. Zhang, arXiv:1502.03928 (2015).

\end{references}
\end{document}